\documentclass[aps,prb,superscriptaddress,showpacs,showkeys]{revtex4}
\usepackage{graphicx}
\newcommand{\be}{\begin{equation}}
\newcommand{\ee}{\end{equation}}
\newcommand{\bea}{\begin{eqnarray}}
\newcommand{\eea}{\end{eqnarray}}
\newcommand{\eps}{\varepsilon}
\newcommand{\ev}[1]{\langle#1\rangle}
\newcommand{\ddt}{\frac{\partial}{\partial t}}
\newcommand{\ihddt}{i\hbar\frac{\partial}{\partial t}}
\newcommand{\mcal}{\mathcal}
\newcommand{\mrm}{\mathrm}

\renewcommand{\vec}[1]{\mathbf{#1}}

\begin{document}
%\preprint{}
%
\title{Influence of Coulomb and Phonon
Interaction on the Exciton Formation 
Dynamics in Semiconductor Heterostructures}
\author{Walter Hoyer}
\affiliation{Department of Physics and Material Sciences Center,
             Philipps University, Renthof 5, D-35032 Marburg, Germany}
\email[]{walter.hoyer@physik.uni-marburg.de}
\author{Mackillo Kira}
\affiliation{Department of Physics and Material Sciences Center,
             Philipps University, Renthof 5, D-35032 Marburg, Germany}
\author{Stephan W. Koch}
\affiliation{Department of Physics and Material Sciences Center,
             Philipps University, Renthof 5, D-35032 Marburg, Germany}
%\thanks{}
\date{\today}
%
%\newpage
%
\begin{abstract}
A microscopic theory is developed to analyze the dynamics of exciton formation\
out of incoherent carriers in semiconductor heterostructures. The carrier
Coulomb and phonon interaction is included consistently.
A cluster expansion method is used to systematically
truncate the hierarchy problem. By including all correlations
up to the four-point (i.e.\ two-particle) level, the fundamental fermionic
substructure of excitons is fully included.  The analysis shows 
that the exciton formation is an intricate process where Coulomb correlations 
rapidly build up on a picosecond time scale while phonon dynamics leads to
true exciton formation on a slow nanosecond time scale.
\end{abstract}
\maketitle
%
%\tableofcontents
%
%%%%%%%%%%%%%%%%%%%%%%%%%%%%%%%%%%%%%%%%%%%%%%%%%%%%%
%
%%%%%%%%%%%%%%%%%%%%%%%%%%%%%%%%%%%%%%%%%%%%%%%%%%%%%%%%%%%%%%%%
%%%%%%%%%%%%%%%%%%%%%%%%%%%%%%%%%%%%%%%%%%%%%%%%%%%%%%%%%%%%%%%%
\section{Introduction}
\label{sec:intro}
%%%%%%%%%%%%%%%%%%%%%%%%%%%%%%%%%%%%%%%%%%%%%%%%%%%%%%%%%%%%%%%%
%%%%%%%%%%%%%%%%%%%%%%%%%%%%%%%%%%%%%%%%%%%%%%%%%%%%%%%%%%%%%%%%
%
The linear absorption spectrum of an
intrinsic direct-gap semiconductor in its ground state
shows excitonic resonances
energetically below the fundamental band-gap energy. 
These resonances in the linear optical polarization
are a consequence of 
the attractive Coulomb interaction between a
conduction-band electron and a valence-band hole.
Mathematically, the eigenvalue problem of an exciton 
in the so-called Wannier limit is
identical to that of the hydrogen atom, such that all
relevant properties are well known.\cite{Haug:94}

However, the appearance of excitonic signatures in
optical spectra does not necessarily imply the population of
excitonic states. Even if the optical excitation is directly
resonant with the exciton energy, the optical excitation always induces
optical polarization in the semiconductor. This polarization,
respectively the absolute square of the polarization, is sometimes
related to {\it coherent populations}; however, it does not
yield information about the truly incoherent population of
single particle or pair states.
 
Furthermore, the optically induced polarization and the
other coherences decay away typically on a picosecond time scale due to:
i) excitation-induced dephasing resulting from the carrier-carrier
Coulomb scattering\cite{Jahnke:96,Jahnke:97},
ii) phonon scattering\cite{Haug:94,Borri:99,Mieck:00}, and 
iii) the finite radiative lifetime\cite{Tassone:92,Andreani:94,Khitrova:99}
in confined semiconductor structures. 
However, after the disappearance of the optical polarization
the material is typically not back in its ground state since
carrier densities and incoherent correlations remain in the
system and continue their many-body dynamics for several nanoseconds.
The life time of these incoherent excitations
is limited only by slow radiative and non-radiative
recombination processes. In ideal semiconductors, only spontaneous recombination
determines the ultimate lifetime of the excited charge carriers.

In this context, an old but still open question
concerns the nature of the incoherent Coulomb correlated populations, in particular
the conditions under which a significant part of the
electron-hole excitations exist in the form of
bound pairs, i.e.\ Wannier excitons.
Although it is not even clear a priori how to count bound states since 
a rigorous exciton number operator does not exist\cite{Usui:60},
one wants to understand the quantum statistical properties of incoherent
excitons, their distribution function, and possible bosonic as well as
Bose condensation aspects.\cite{Snoke:87,Johnsen:01}

So far, several papers discuss the relaxation dynamics of excitons
due to phonons together with a direct
bosonic approximation for the exciton\cite{Gulia:97,Ivanov:99}. In other approaches,
exciton formation after coherent excitation has been investigated,
however, restricting the treatment to weak excitation conditions
due to a low-order expansion in terms of
the exciting pulse.\cite{Thraen:00,Siantidis:01} Since these
approaches do not involve fermionic carrier densities and carrier-carrier
correlations, they cannot resolve how much of the excitation is transferred
directly into those correlations and densities. Typically, in bosonic models,
exciton contributions are separated into bound and continuum excitons. In this
description, one cannot determine the plasma carrier density since the
continuum excitons determined from the two-particle reduced density matrix
are not trivially related to the carrier densities given by the one-particle
reduced density matrix.

In this paper, we present and utilize a microscopic theory for the description of
the totally incoherent regime. In this approach, fermionic carriers, incoherent
exciton populations, carrier-carrier correlations, as well as phonon-assisted
correlations are treated at the same level. As a result, we are
able to evaluate the excitonic correlations, formation rates,
distribution functions, etc.~for different temperatures and carrier densities
without neglecting the underlying fermionic anti-symmetry. 

After presenting the theoretical background and the full equations in the
incoherent limit in Sec.~\ref{sec:theory}, we discuss our numerical
results in Sec.~\ref{sec:exform}. This is done by
viewing the build-up of correlations and monitoring the growth of the
electron-hole pair-correlation function. We investigate the issue of energy
conservation and derive an approximate adiabatic solution which offers the 
possibility to obtain an intuitive interpretation of the exciton formation
process. The results are summarized in Sec.~\ref{sec:summary} and
some technical details of the calculations are presented in the appendices.
%
%%%%%%%%%%%%%%%%%%%%%%%%%%%%%%%%%%%%%%%%%%%%%%%%%%%%%%%%%%%%%%%%%%%%%%%%%%%
%%%%%%%%%%%%%%%%%%%%%%%%%%%%%%%%%%%%%%%%%%%%%%%%%%%%%%%%%%%%%%%%%%%%%%%%%%%
%
%%%%%%%%%%%%%%%%%%%%%%%%%%%%%%%%%%%%%%%%%%%%%%%
%%%%%%%%%%%%%%%%%%%%%%%%%%%%%%%%%%%%%%%%%%%%%%%%%
\section{Incoherent Exciton and Plasma Dynamics}
\label{sec:theory}
%%%%%%%%%%%%%%%%%%%%%%%%%%%%%%%%%%%%%%%%%%%%%%%%%
%%%%%%%%%%%%%%%%%%%%%%%%%%%%%%%%%%%%%%%%%%%%%%%%%
%

In contrast to earlier investigations of exciton formation
after resonant excitation\cite{Thraen:00,Siantidis:01}, we
focus in this paper on an initially completely incoherent
configuration. Such a situation can be realized experimentally
either by pulsed continuum excitation, after the initial
coherences have decayed, of with a current injection of carriers since
this process does not induce optical coherences. In all numerical investigations in
this paper, we therefore start with vanishing correlations and finite carrier
densities and view the evolution of the system developing from that initial
condition.

%%%%%%%%%%%%%%%%%%%%%%%%%%%%%%%
%%%%%%%%%%%%%%%%%%%%%%%%%%%%%%%
\subsection{The Hamiltonian}
\label{sec:hamiltonian}
%%%%%%%%%%%%%%%%%%%%%%%%%%%%%%%
%%%%%%%%%%%%%%%%%%%%%%%%%%%%%%%
%
The starting point of our theoretical description is the full Hamiltonian which
describes the free motion of non-interacting carriers and phonons as 
well as the interaction between those quasi-particles.
In this section, we briefly outline the Hamiltonian which is used throughout the 
paper; for more detailed derivations, see e.g.\ Refs.~{\cite{Haug:94,Kira:99}.

In semiconductor heterostructures, the microscopic proper{\-}ties of the
carriers can be described with the fermionic field operator ${a}_{\lambda, k}$
(${a}^{\dagger}_{\lambda, k}$) annihilating (creating) an electron in
band~$\lambda$ with wave vector $k$ along the heterostructure.
Here, in general, the band index $\lambda$ may include different bands, subbands,
and spins. For sufficiently narrow confinement, the carrier dynamics is
restricted practically completely to the lowest subband and the Bloch functions
follow from the envelope approximation.\cite{Haug:94}
The corresponding free Hamiltonian is
%
%%%%%%%%%%%%%%%%%
\be
{H}_{\mrm{kin}} =
\sum_{k} 
\left(
\eps^{{c}}_{k}\, a^{\dagger}_{c,k} a_{c,k}
+
\eps^{{v}}_{k}\, a^{\dagger}_{v,k} a_{v,k}
\right),
\label{eq:H0kin}
\ee
%%%%%%%%%%%%%%%%%
%
which includes one valence and one
conduction band. The generalization to a multi-band system or the inclusion of
spin is straight forward.\cite{Haug:94,Hader:98} In general, the eigenenergies
$\eps^{c/v}_{k}$ must be obtained from precise band structure calculations.
For the investigation of near-bandgap optical features, we may use
$\eps^{c}_{k} \equiv \eps^{e}_{k} + E_{\mrm{G}}
= \frac{\hbar^2 k^2}{2 m^{e}} + E_{\mrm{G}}$ and
$\eps^{v}_{k} \equiv -\eps^{h}_{k} = - \frac{\hbar^2 k^2}{2 m^{h}}$
where $m^e$ and $m^{h}$ are the effective electron and hole masses. The unrenormalized
bandgap is denoted with $E_{\mrm{G}}$. 

In order to include the Coulomb interaction between the electrons in different bands,
we use the Hamiltonian\cite{Haug:94,Jauho:96}
%
%%%%%%%%%%%%%%%%%
\be
{H}_{\mrm{C}}
=
\frac{1}{2}\sum_{k,k',q\not=0} V_q 
\left(
a^{\dagger}_{c,k} a^{\dagger}_{c,k'} a_{c,k'+q} a_{c,k-q}
+
a^{\dagger}_{v,k} a^{\dagger}_{v,k'} a_{v,k'+q} a_{v,k-q}
+
2\, a^{\dagger}_{c,k} a^{\dagger}_{v,k'} a_{v,k'+q} a_{c,k-q}
\right),
\label{eq:HC}
\ee
%%%%%%%%%%%%%%%%%
%
where $V_q$ is the quantum-well or quantum-wire Coulomb matrix element. The first
two terms in Eq.~(\ref{eq:HC}) lead to repulsive interaction between electrons
within the same band whereas the last term gives rise to the attractive interaction
between electrons and holes (missing valence band electrons) in different bands.

In semiconductors, the carriers are additionally coupled to lattice vibrations (phonons).
The noninteracting part of the phonon Hamiltonian is
%
%%%%%%%%%%%%%%%%%
\be
{H}_{\mrm{phon}} =
\sum_{\bf p}
\hbar \Omega_{|{\bf p}|} \left( D^{\dagger}_{\bf p} D_{\bf p} + \frac{1}{2} \right),
\label{eq:H0phon}
\ee
%%%%%%%%%%%%%%%%%%
%
where the bosonic operator $D_{\bf p}$ ($D^{\dagger}_{\bf p}$) annihilates (creates) a
phonon in the state~${\bf p}$. Here, ${\bf p}=(p,p_{\perp},\sigma)$ labels the
wave vector along and perpendicular to the heterostructure as well as the phonon
branch. Corresponding to the three independent modes of sound waves in a solid,
three branches of acoustic phonons always exist in a three-dimensional
semiconductor.\cite{Ashcroft:76} In lattices with more than one atom within a
unit cell, additional optical phonon branches are present.

The carrier-phonon interaction Hamiltonian can be expressed in the form
%
%%%%%%%%%%%%%%%%%%%%
%%%%%%%%%%%%%%%%%%%%
\begin{equation}
{H}_{\mrm{P}} = \sum_{p,p_{\perp},k}
        D^{\dagger}_{p,p_{\perp}}
        \left(
        G^{{c}}_{p,p_{\perp}}
        a^{\dagger}_{c,k} a_{c,k+p} 
        +
        G^{{v}}_{p,p_{\perp}}
        a^{\dagger}_{v,k} a_{v,k+p} 
        \right)
        \mbox{ + h.c.}\,,
\label{eq:HP}
\end{equation}
%%%%%%%%%%%%%%%%%%%%
%%%%%%%%%%%%%%%%%%%%
%
where $G^{\lambda}_{p,p_{\perp}}$ is the effective coupling matrix-element between
a phonon and an electron in band $\lambda$.
Only the momentum component $p$ along the heterostructure is conserved because
the Bloch electrons are confined in perpendicular direction $p_{\perp}$. The form
of Eq.~(\ref{eq:HP}) reflects the fact that phonon interaction takes place
within one band because phonons cannot provide the energy for an interband transition.
The exact form of the interaction matrix element $G^{{c(v)}}_{p,p_\perp}$
depends on the interaction type which is included (i.e. deformation potential,
piezo-electric coupling, etc.). For the treatment of acoustic phonons in the
long-wavelength limit, the matrix element is given by a product between the
plain matrix element 
$C_{\vec{p}} = \sqrt{\frac{\hbar |D|^2 |\vec{p}|}{2 V c_{\mrm{LA}} \rho}}$,
where  $\rho$ denotes the mass density and $V$ the volumne of the semiconductor
material, $c_{\mrm{LA}}$ is the velocity of sound and $D$ the deformation constant,
and the form factor $f_R(p_{\perp})$ which is a consequence of the confinement
of the electrons in the direction perpendicular to the semiconductor structure.
By choosing parabolic confinement, we obtain
$f_R(p_{\perp}) = e^{-\frac{p^2_{\perp} R^2}{2}}$
for a one-dimensional quantum wire characterized by the confinement potential
$\frac{\hbar^2 r_{\perp}^2}{2 m R^4}$.

The starting point of all our further investigations is the total 
Hamiltonian of Eq.~(\ref{eq:H0kin})--(\ref{eq:HP}),
%
%%%%%%%%%%%%%%%%
\be
{H}_{\mrm{tot}} =
{H}_{\mrm{kin}} +
{H}_{\mrm{C}} +
{H}_{\mrm{phon}} + 
{H}_{\mrm{P}},
\label{eq:Htot}
\ee
%%%%%%%%%%%%%%%%
%
where both interaction parts of the total Hamiltonian lead to non-trivial 
coupling to higher order correlations.
%The contributions of the non-interacting carriers and phonons will be lumped together into ${H}_{0}$.

%
%%%%%%%%%%%%%%%%%%%%%%%
\subsection{Hierarchy Problem}
\label{sec:hierarchy}
%%%%%%%%%%%%%%%%%%%%%%%
%
Starting from Eq.~(\ref{eq:Htot}), one can use the Heisenberg equation to
compute the equations of motion for all relevant expectation values of interest.
The simplest examples of incoherent one-particle expectation values are the
two-point quantities describing the electron and hole distributions
%
%%%%%%%%%%%%%%%%%%%%
\be
f^{e}_{k} = \langle a^{\dagger}_{c,k} a_{c,k}\rangle,\quad
f^{h}_{k} = \langle a_{v,-k} a^{\dagger}_{v,-k}\rangle\,.
\label{eq:def_f}
\ee
%%%%%%%%%%%%%%%%%
%
As is well known, the dynamics of $f^{\lambda}_{k}$ does
not yield a closed set of equations since
both the Coulomb and phonon interaction Hamiltonians lead to coupling of
the two-point quantities to either four-point quantities or mixed carrier-phonon
operators. Schematically, the hierarchy problem\cite{Haug:94,Schaefer:02} is described by
%
%%%%%%%%%%%%%%%%%
\be
\ihddt \ev{N} = T[\ev{N}] + V[\ev{N+1}]
\label{eq:hierarchy}
\ee
%%%%%%%%%%%%%%%%%
%
where an $N$-particle (i.e.\ 2$N$-point) operator is coupled to $(N+1)$-particle
operators due to the many-body interaction.
To solve this hierarchy problem rigorously, we apply a method known from quantum
chemistry where the truncation problem for many-electron wave functions has successfully
been approached with the so-called cluster expansion.\cite{Cizek:66,Purvis:82,Harris:92}
There, the electronic wave function is divided into classes where electrons are:
i) independent single particles (singlets), ii) coupled in pairs (doublets), iii) coupled
in triplets, and iv) coupled in higher order clusters. The $N$-particle wave function is
then approximated by a suitable amount of coupled clusters by including the correct
antisymmetry of the fermions.

In semiconductors, the system properties can be
evaluated from $2N$-point expectation values
$ \ev{N}
  \equiv
  \ev{
  a_{\lambda_1,k_1}^{\dagger} .. a_{\lambda_N,k_N}^{\dagger}
  a_{\nu_N,p_N}  .. a_{\nu_1,p_1}
  }$,
which determine the reduced density matrix in the Bloch basis. 
If the system contains exactly ${\mcal N}$ particles, $\ev{\mcal{N}}$
fully describes the system properties. In this case, $2N$-point expectation
values exist for all $N \le {\cal N}$.
The simplest approximation scheme is provided by the Hartree-Fock
approximation which implicitly assumes that the many-body system is
described by a so-called Slater determinant of ${\cal N}$
independent single-particle wave functions. In this case,
$2N$-point expectation values can be expressed in terms of
two-point expectation values. For example, a four-point expectation
value in singlet (i.e.~Hartree-Fock) approximation is given by
%
%%%%%%%%%%%%%%%%%
\be
\ev{
  a_{\lambda_1,k_1}^{\dagger} a_{\lambda_2,k_2}^{\dagger}
  a_{\nu_2,p_2} a_{\nu_1,p_1}
}_{\mrm{S}}
=
\ev{a_{\lambda_1,k_1}^{\dagger} a_{\nu_1,p_1}}
\ev{a_{\lambda_2,k_2}^{\dagger} a_{\nu_2,p_2}}
-
\ev{a_{\lambda_1,k_1}^{\dagger} a_{\nu_2,p_2}}
\ev{a_{\lambda_2,k_2}^{\dagger} a_{\nu_1,p_1}}.
\ee
%%%%%%%%%%%%%%%%%
%
In the spirit of the original cluster expansion
the system is thus fully described by uncorrelated singlets
where each carrier behaves effectively like a single particle
influenced by the meanfield of all other particles.
Higher order contributions are defined recursively. For example, true
two-particle correlations are obtained from
%
%%%%%%%%%%%%%%%%%
\be
\Delta\ev{
  a_{\lambda_1,k_1}^{\dagger} a_{\lambda_2,k_2}^{\dagger}
  a_{\nu_2,p_2} a_{\nu_1,p_1}
}
=
\ev{
  a_{\lambda_1,k_1}^{\dagger} a_{\lambda_2,k_2}^{\dagger}
  a_{\nu_2,p_2} a_{\nu_1,p_1}
}
-
\ev{
  a_{\lambda_1,k_1}^{\dagger} a_{\lambda_2,k_2}^{\dagger}
  a_{\nu_2,p_2} a_{\nu_1,p_1}
}_{\mrm{S}},
\ee
%%%%%%%%%%%%%%%%%
%
i.e., by subtracting the singlet contribution from the full expectation value.
The advantage of this factorization is its direct physical interpretation. By subtracting
the single-particle contribution, the resulting
correlated part really describes the true two-particle correlations.\cite{Hoyer:02d,Kira:02a}
In the same way, truly correlated triplets and higher clusters can be defined by
subtracting all lower-level contributions.\cite{Fricke:96}

If we only include plasma and pair-correlation
effects in the analysis, an arbitrary $N$-particle 
expectation value can be expressed consistently
with  the singlet-doublet approximation 
$\ev{N} \approx \ev{N}_{\mrm{SD}}$.
This determines uniquely how the truncation of the hierarchy has to be
performed; we only need to --- and are allowed to --- solve dynamics of
$\ev{1}$ and $\Delta\ev{2}$ because any arbitrary $\ev{N}_{\mrm{SD}}$
consists only of two-point expectation values and four-point
correlations. According to Eq.~(\ref{eq:hierarchy}), 
we obtain
%-----------------------
%-----------------------
\begin{eqnarray}
\ihddt \ev{1} & = & T_1[\ev{1}] + V_1[\ev{2}_{\mrm{S}}] + V_1[\Delta\ev{2}],
\label{eq:truncHEM1},
\\
\ihddt \Delta\ev{2} & = & T_2[\Delta\ev{2}] + V_2[\ev{3}_{\mrm{SD}}]
\label{eq:truncHEM2},
\end{eqnarray}
%-----------------------
%-----------------------
where $T_{1(2)}$ and $V_{1(2)}$ are known functionals defined by the
specific form of the Heisenberg equation of motion. 
The consistent singlet-doublet-approximation is obtained when
$\ev{3}$ is approximated by $\ev{3}_{\mrm{SD}}$; as a result, the
infinite hierarchy is systematically truncated and
Eqs.~(\ref{eq:truncHEM1})--(\ref{eq:truncHEM2}) are closed.
In order to study exciton formation, we consequently
have to know the exciton and carrier-carrier correlations
%
%%%%%%%%%%%%%%%%%%%%
\bea
c_{{X}}^{q,k',k}
        & = &
\Delta \langle a^{\dagger}_{c,k} a^{\dagger}_{v,k'} a_{c,k'+q} a_{v,k-q} \rangle,
\label{eq:def_cX}
\\
c_{e}^{q,k',k}
        & = &
\Delta \langle a^{\dagger}_{c,k} a^{\dagger}_{c,k'} a_{c,k'+q} a_{c,k-q} \rangle,
\label{eq:def_ce}
\\
c_{h}^{q,k',k}
        & = &
\Delta \langle a^{\dagger}_{v,k} a^{\dagger}_{v,k'} a_{v,k'+q} a_{v,k-q} \rangle,
\label{eq:def_ch}
\eea
%%%%%%%%%%%%%%%%%%%%
%
in addition to the single-particle densities, Eq.~(\ref{eq:def_f}).

In general, the $\ev{N}_{\rm SD} $ truncation
fully includes the $\chi^{(3)}$ limit of the so-called 
dynamics controlled truncation scheme.\cite{Lindberg:94,Axt:94a}
Obviously, a similar relation holds
as the number of clusters is increased; i.e.~$\chi^{(2N-1)}$ is
a subset of the $N$-particle cluster expansion. 
Furthermore, the cluster expansion can be used in regimes where
$\chi^{(N)}$ methods fail, e.g., when the exciting field is strong
or the system becomes fully incoherent.
When $\ev{3}$ is approximated by including only the singlet part
$\ev{3}_{\mrm{S}}$ in Eq.(\ref{eq:truncHEM2}), also the second Born
approach\cite{Jahnke:97} is found to be a subset of the cluster expansion.
Since in this case the doublet part is included only partially,
the second Born scheme does 
not allow formation of bound states but rather describes
microscopic scattering. This ideology can be directly 
generalized in order to include the triplets in the scattering level; 
for such an approach, the $\Delta \ev{3}$ dynamics is solved but
the resulting $\ev{4}_{\mrm{SDT}}$ is approximated by $\ev{4}_{\mrm{SD}}$.
This simplifies the numerical complexity of the triplet terms
considerably and allows their analytic evaluation 
on the scattering level.

The cluster expansion can also be used directly to classify and truncate
carrier-phonon correlations using the formal equivalence between a
boson operator and a pair of fermion operators.
The operator equation of motion,
%
%%%%%%%%%%%%%%%%%%%%
\be
\left[
\ihddt D_{p,p_\perp}
\right]_{H_{\mrm{phon}}+H_{\mrm{P}}}
=
\hbar\Omega_{p,p_\perp}\,  D_{p,p_\perp}
+
\sum_{\lambda,k} \mcal{G}^{\lambda}_{p,p_\perp} a^{\dagger}_{\lambda,k} a_{\lambda,k+q},
\label{eq:formalD}
\ee
%%%%%%%%%%%%%%%%%%%%
%
shows that one phonon operator is formally equivalent to a product of two electronic
operators and that with every phonon absorption or emission process, an electron
changes its momentum. In that sense, also the phonon interaction leads to an
infinite hierarchy formally equivalent to the many-body hierarchy problem discussed above.
For example,
$ \Delta \ev{ D_{p,p_{\perp}} a_{\lambda,k}^{\dagger} a_{\lambda,k-p} }
= 
\ev{ D_{p,p_{\perp}} a_{\lambda,k}^{\dagger} a_{\lambda,k-p} }
-
\ev{ D_{p,p_{\perp}} } \ev{ a_{\lambda,k}^{\dagger} a_{\lambda,k-p} } $
describes correlations between carriers and phonons.
A similar treatment of the quantized light-matter
interaction has led to quantum-optical effects resulting, e.g., in
squeezing and entanglement.\cite{Kira:99,Lee:99,Ell:00} The singlet-doublet
truncation with a controlled scattering treatment of the triplets leads to a closed
set of equations which includes the dominant
correlations between lattice vibrations and carriers.

%
%%%%%%%%%%%%%%%%%%%%%%%%%%%%%%%%%%%
\subsection{Coulomb Interaction}
\label{sec:eomHC}
%%%%%%%%%%%%%%%%%%%%%%%%%%%%%%%%%%%
%
Since the Heisenberg equations of motion are linear with respect to the different
parts of the total Hamiltonian (\ref{eq:Htot}), we examine these different parts
separately.
For the exciton correlations under the influence of Coulomb interaction, we obtain
%%%%%%%%%%%%%%%%%%%%
%%%%%%%%%%%%%%%%%%%%
\begin{eqnarray}
\lefteqn{
  \left[
  i \hbar\ddt
  c_{{X}}^{q,k',k} 
  \right]_{H_{\mrm{kin}}+H_{\mrm{C}}}
=
\left(
    \tilde{\epsilon}^{e}_{k'+q} 
    + 
    \tilde{\epsilon}^{h}_{k'}
    -
    \tilde{\epsilon}^{h}_{k-q}
    - 
    \tilde{\epsilon}^{e}_{k}
  \right)
  c_{{X}}^{q,k',k}
}
\nonumber\\
 &&
 + V_{k-k'-q} 
  \left[ 
    \left(1- f^{e}_{k}\right)
    \left(1-f^{h}_{k-q}\right) 
    f^{e}_{k'+q}
    f^{h}_{k'}
-
    f^{e}_{k}
    f^{h}_{k-q}
    \left(1-f^{e}_{k'+q}\right)
    \left(1-f^{h}_{k'}\right)
  \right]
\nonumber\\
   &&
   +V_{k-k'-q}
   \left[
     \left(
       f^{h}_{k-q}-f^{h}_{k'}
     \right)
     \sum_{l}
     c_{{e+X}}^{q+k'-l,l,k}
     +
     \left(
       f^{e}_{k}-f^{e}_{k'+q}
     \right)
     \sum_{l}
     c_{{h+X}}^{q-k+l,k',l}
     \right]
\nonumber\\
&&
   +\left[
     1-f^{e}_{k}-f^{h}_{k-q}
   \right]
   \sum_{l} 
   V_{l-k}
   c_{{X}}^{q,k',l}
  -\left[
     1-f^{e}_{k'+q}-f^{h}_{k'}
   \right]
   \sum_{l} 
   V_{l-k'}
   c_{{X}}^{q,l,k}
\nonumber\\
&&
   +\left[
     f^{h}_{k-q}-f^{e}_{k'+q}
   \right]
   \sum_{l} 
   V_{l-q}
   c_{{X}}^{l,k',k}
+
   \left[
     f^{e}_{k}-f^{h}_{k'}
   \right]
   \sum_{l}
   V_{l-q}
   c_{{X}}^{l,q+k'-l,k-q+l}
\nonumber\\
&&
+
   \left[
     f^{e}_{k'+q}-f^{e}_{k}
   \right]
   \sum_{l} 
   V_{l-k}
   c_{{X}}^{q-k+l,k',l}
+
   \left[
     f^{h}_{k'}-f^{h}_{k-q}
   \right]
   \sum_{l} 
   V_{l-k'}
   c_{{X}}^{q+k'-l,l,k}
\label{eq:cvcv},
\end{eqnarray}
%%%%%%%%%%%%%%%%%%%%
%%%%%%%%%%%%%%%%%%%%
which is coupled both to carrier densities and to electron and hole correlations 
$c_{{e(h)}}^{q,k',k}$ via the terms $c_{e(h)+X}^{q,k',k} = c_{e(h)}^{q,k',k}
+ c_{X}^{q,k',k}$. The dynamics of these correlations is described by similar
equations given in App.~\ref{app:carrcorr}.

The interpretation of Eq.~(\ref{eq:cvcv}) is straightforward; the first line
gives the kinetic evolution of the four-point operator with the renormalized energies 
$\tilde{\epsilon}^{e(h)}_k = \epsilon^{e(h)}_k - \sum_{k'} V_{k-k'} f^{e(h)}_{k'}$.
The second line contains the factorized source term which initiates
the creation of excitonic correlations as soon as electrons and holes are present. This
singlet source is altered by the direct influence of the correlations in the third line.
The remaining six Coulomb sums describe the different possibilities how two out of the four
fermions in $\Delta\ev{a^{\dagger}_c a^{\dagger}_v a_c a_v}$ can interact via the Coulomb
interaction. These sums lead to the possibility to
form bound excitons. The major contributions originate from the first two of the six
sums which are multiplied by a phase space filling factor instead of a density
difference and are thus appreciable even for low densities.
All other sums vanish for low density but become important when the density is
increased. They are a consequence of the indistinguishability of electrons and holes
and correspond to Coulomb interaction between carriers formally ``belonging'' to two
different excitons. Equations investigating exciton formation in the
$\chi^{(3)}$-limit\cite{Thraen:00,Siantidis:01} do neither contain these additional
$f$-dependent contributions nor the singlet source.

In order to obtain a closed set of equations for the incoherent dynamics of the
pure carrier system, we also have to derive the equation of motion for electron
and hole densities. They are given by
%
%%%%%%%%%%%%%%%%%%%%
\bea
\left[
   \ddt
   f^{e}_{k}
\right]_{H_\mrm{kin}+H_\mrm{C}}
& = &
   \frac{2}{\hbar} \mrm{Im}
        \left[
  \sum_{k',q} V_{q} c_{e}^{q,k',k}
- \sum_{k',q} V_{k-q-k'} c_{X}^{q,k',k} 
        \right],
\label{eq:SBEcc}
\\
\left[
   \ddt
   f^{h}_{k}
\right]_{H_\mrm{kin}+H_\mrm{C}}
& = &
   \frac{2}{\hbar} \mrm{Im}
        \left[
- \sum_{k',q} V_{q} c_{h}^{q,k',k}
+ \sum_{k',q} V_{k-q-k'} c_{X}^{-q,k,k'}
        \right]
\label{eq:SBEvv}
\eea
%%%%%%%%%%%%%%%%%%%%
%
and describe the influence of the correlations on the carrier distributions. Since all incoherent
four-point correlations are treated exactly, this approach includes the microscopic carrier-carrier
scattering beyond the second Born approach.\cite{Jahnke:97}

%
%%%%%%%%%%%%%%%%%%%%%%%%%%%%%%%%%%%
\subsection{Phonon Interaction}
\label{sec:eomHP}
%%%%%%%%%%%%%%%%%%%%%%%%%%%%%%%%%%%
%
As a consequence of the Coulomb interaction, excitonic
correlations can build up since the carriers of opposite charge are attracted towards
each other. However, two
particles cannot be bound without the assistance of a third object due to
energy and momentum conservation. In a many-body system, any additional
carrier can be this third object and compensate the energy gained by the 
bound-state formation. This
process inevitably leads to heating of the remaining carrier system such that further
formation gets more and more improbable. By a coupling 
to phonons the excess energy can be directed out of the many-body carrier
system into the phonon reservoir. For simplicity, we assume that
the phonon system is basically unperturbed
by the carrier dynamics and solve the phonon dynamics using the
steady state Markov approximation with the assumption of thermal occupation of the
different phonon states. Working out the phonon contributions to the
dynamic equations, we obtain in the singlet-doublet approximation
%
%%%%%%%%%%%%%%%%%%%%
\begin{eqnarray}
\left[
\ddt f^{e}_{k}
\right]_{H_{\mrm{P}}}
& = & \frac{2}{\hbar}\sum_{p} 
\mrm{Im}\left[\Pi^{e}_{k,p}\right]\,,
\label{eq:ddtfephon2}
\\
\left[
\ddt f^{h}_{k}
\right]_{H_{\mrm{P}}}
& = & -\frac{2}{\hbar}\sum_{p} 
\mrm{Im}\left[\Pi^{h}_{k,p}\right]\,,
\label{eq:ddtfhphon2}
\\
%%%%%%%%%%%%%%%
\left[  
i \hbar\ddt
  c_{{X}}^{q,k',k}
\right]_{H_{\mrm{P}},\mrm{SD}}
&=&
\Pi^{e}_{k,k-q-k'}\left(f^{h}_{k'}-f^{h}_{k-q}\right)
-\,
\Pi^{h}_{k',k'+q-k}\left(f^{e}_{k}-f^{e}_{k'+q}\right)\,,
\nonumber\\
\label{eq:ddtcXphon4p}
\\
\left[
  i \hbar\ddt
  c_{e}^{q,k',k}
\right]_{H_{\mrm{P}},\mrm{SD}}
&=&
\Pi^{e}_{k,k-q-k'}\left(f^{e}_{k-q}-f^{e}_{k'}\right)
+
\Pi^{e}_{k',k'+q-k}\left(f^{e}_{k'+q}-f^{e}_{k}\right)
\nonumber\\
&&
+\,
\Pi^{e}_{k',-q}\left(f^{e}_{k}-f^{e}_{k-q}\right)
+
\Pi^{e}_{k,q}\left(f^{e}_{k'}-f^{e}_{k'+q}\right),
\label{eq:ddtcephon4p}
\\
%%%%%%%%%%%%%%%%%%
\left[
  i \hbar\ddt
  c_{h}^{q,k',k}
\right]_{H_{\mrm{P}},\mrm{SD}}
&=&
\Pi^{h}_{k,k-q-k'}\left(f^{h}_{k'}-f^{h}_{k-q}\right)
+
\Pi^{h}_{k',k'+q-k}\left(f^{h}_{k}-f^{h}_{k'+q}\right)\,,
\nonumber\\
&&
+\,
\Pi^{h}_{k',-q}\left(f^{h}_{k-q}-f^{h}_{k}\right)
+
\Pi^{h}_{k,q}\left(f^{h}_{k'+q}-f^{h}_{k'}\right)
\label{eq:ddtchphon4p}
\end{eqnarray}
%%%%%%%%%%%%%%%%%%%%
%
where we have defined the term
%
%%%%%%%%%%%%%%%%%%%%
\begin{eqnarray}
\Pi^{e}_{k,p}
&=&
   i \sum_{p_\perp} |G_{p,p_\perp}|^2 
        f^{e}_{k-p} (1-f^{e}_{k})
        \Bigl\{
                N_{p,p_\perp}^{\mrm{PH}}
        g(\eps^{e}_{k}-\eps^{e}_{k-p}-\hbar\Omega_{p,p_\perp})
\nonumber\\
&& \qquad\qquad \qquad\qquad\qquad
                + (N_{p,p_\perp}^{\mrm{PH}} +1) 
        g(\eps^{e}_{k}-\eps^{e}_{k-p}+\hbar\Omega_{p,p_\perp})
        \Bigr\}
\nonumber\\
&&
  -i \sum_{p_\perp} |G_{p,p_\perp}|^2 
        f^{e}_{k} (1-f^{e}_{k-p})
        \Bigl\{
                N_{p,p_\perp}^{\mrm{PH}}
        g(\eps^{e}_{k}-\eps^{e}_{k-p}+\hbar\Omega_{p,p_\perp})
\nonumber\\
&& \qquad\qquad\qquad\qquad\qquad
                + (N_{p,p_\perp}^{\mrm{PH}} +1) 
        g(\eps^{e}_{k}-\eps^{e}_{k-p}-\hbar\Omega_{p,p_\perp})
        \Bigr\}
\nonumber\\
&&
+i \sum_{l,p_\perp} |G_{p,p_\perp}|^2 
        c_e^{p,l,k}
\nonumber\\
&&\qquad\quad
        \Bigl\{
        g(\eps^{e}_{l+p}-\eps^{e}_{l}+\hbar\Omega_{p,p_\perp})
        -
        g(\eps^{e}_{l+p}-\eps^{e}_{l}-\hbar\Omega_{p,p_\perp})
        \Bigr\},
\nonumber\\
&&
-i \sum_{l,p_\perp} |G_{p,p_\perp}|^2
        c_X^{k-p-l,l,k}
\nonumber\\
&&\qquad\quad
        \Bigl\{
        g(\eps^{h}_{l}-\eps^{h}_{l+p}+\hbar\Omega_{p,p_\perp})
        -
        g(\eps^{h}_{l}-\eps^{h}_{l+p}-\hbar\Omega_{p,p_\perp})
        \Bigr\},
\label{eq:definePHe}
\end{eqnarray}
%%%%%%%%%%%%%%%%%%%%
%
and a similar expression for $\Pi^{h}_{k,p}$
with $g(\eps)=\pi\delta(\eps) + i P(\frac{1}{\eps})$. The derivation of
Eqs.~(\ref{eq:ddtfephon2})--(\ref{eq:definePHe}) is presented in App.~\ref{app:phonons}.
After being inserted into Eqs.~(\ref{eq:ddtfephon2}) and~(\ref{eq:ddtfhphon2}), the first
four lines of Eq.~(\ref{eq:definePHe}) have a straight forward
interpretation as scattering rates. They are the
typical scattering equations with a balance between scattering into (lines one and two)
and scattering out of (lines three and four) electron state $k$.
In both cases, phonon absorption and emission processes are possible. Since
emission can happen even without any phonons present, the respective terms are
proportional to $N^{\mrm{PH}}_{p,p_\perp}+1$. The last two lines are due to
phonon coupling to higher order correlations. Since both the carrier and the
correlation dynamics in Eqs.~(\ref{eq:ddtfephon2})--(\ref{eq:ddtchphon4p}) are
determined by similar $\Pi$-terms, the corresponding phonon
effects result from the cooling of the carrier plasma. 

The actual exciton formation dynamics is described in the six-point (triplet) level.
The major contribution of these triplet correlations can be evaluated at the scattering
level\cite{Kira:02a} in analogy to the second-Born approximation for four-point
correlations which has successfully been used to describe microscopic Coulomb and
phonon scattering.\cite{Haug:94,Jahnke:97}
Physically, this approach provides scattering of excitons with a 
third object excluding the formation of bound three-particle clusters.
As discussed in App.~\ref{app:phonons}, the resulting equations can be written as
%
%%%%%%%%%%%%%%%%%%%%
%%%%%%%%%%%%%%%%%%%%
\begin{eqnarray}
\left[
  i \hbar\ddt
  c_{{X}}^{q,k',k}
\right]_{H_{\mrm{P}},\mrm{T}}
&=&
\sum_{p}
\left(
\tilde{\gamma}^{h}_{k-q,p} + \tilde{\gamma}^{e}_{k'+q,p}
-
(\tilde{\gamma}^{h}_{k',p})^* - (\tilde{\gamma}^{e}_{k,p})^*
\right)
c_X^{q,k',k}
\nonumber\\
&&
-\,
\sum_{p}
\left(
\gamma^{e}_{k,k-p} - (\gamma^{h}_{k-q,k-p})^*
\right)
c_X^{q,k',p}
\nonumber\\
&&
-
\sum_{p}
\left(
\gamma^{h}_{k',k'-p} - (\gamma^{e}_{k'+q,k'-p})^*
\right)
c_X^{q,p,k}
\nonumber\\
&&
-
\sum_{p}
\left(
(\gamma^{h}_{k-q,p-q})^* + (\gamma^{e}_{k'+q,q-p})^*
\right)
c_X^{p,k',k}
\nonumber\\
&&
+
\sum_{p}
\left(
\gamma^{h}_{k',p-q} + \gamma^{e}_{k,q-p}
\right)
(c_X^{p,k-q,k'+q})^*
\nonumber\\
&&
-
\sum_{p}
\left(
\gamma^{h}_{k',p-q} - (\gamma^{h}_{k-q,p-q})^*
\right)
c_X^{p,k'+q-p,k}
\nonumber\\
&&
-
\sum_{p}
\left(
\gamma^{e}_{k,q-p} - (\gamma^{e}_{k'+q,q-p})^*
\right)
c_X^{p,k',k-q+p}\,,
\label{eq:ddtcXphon6p}
\end{eqnarray}
%%%%%%%%%%%%%%%%%%%%
%%%%%%%%%%%%%%%%%%%%
%
where $\gamma$ and $\tilde{\gamma}$ are defined according to
%
%%%%%%%%%%%%%%%%%%%%
%%%%%%%%%%%%%%%%%%%%
\begin{eqnarray}
\gamma^{e}_{k,p} & = & 
           (-i) \sum_{p_\perp} |G_{p,p_\perp}|^2
           \left[
             (N^{\mrm{PH}}_{p,p_\perp}+1-f^{e}_{k}) 
             g(\eps^{e}_{k}-\eps^{e}_{k-p} + \hbar\Omega_{p,p_\perp})
           \right.
\nonumber\\
&&\qquad\qquad
+
           \left.
             (N^{\mrm{PH}}_{p,p_\perp}+f^{e}_{k}) 
             g(\eps^{e}_{k}-\eps^{e}_{k-p} - \hbar\Omega_{p,p_\perp})
           \right]\,,
\label{eq:definegammae}
\\
\gamma^{h}_{k,p} & = & 
           (-i) \sum_{p_\perp} |G_{p,p_\perp}|^2
           \left[
             (N^{\mrm{PH}}_{p,p_\perp}+f^{h}_{k}) 
             g(\eps^{h}_{k-p}-\eps^{h}_{k} + \hbar\Omega_{p,p_\perp})
           \right.
\nonumber\\
&&\qquad\qquad
+
           \left.
             (N^{\mrm{PH}}_{p,p_\perp}+1-f^{h}_{k}) 
             g(\eps^{h}_{k-p}-\eps^{h}_{k} - \hbar\Omega_{p,p_\perp})
           \right]\,,
\label{eq:definegammah}
\\
\tilde{\gamma}^{e}_{k,p} & = & 
           (-i) \sum_{p_\perp} |G_{p,p_\perp}|^2
           \left[
             (N^{\mrm{PH}}_{p,p_\perp}+f^{e}_{k-p}) 
             g(\eps^{e}_{k}-\eps^{e}_{k-p} + \hbar\Omega_{p,p_\perp})
           \right.
\nonumber\\
&&\qquad\qquad
+
           \left.
             (N^{\mrm{PH}}_{p,p_\perp}+1-f^{e}_{k-p}) 
             g(\eps^{e}_{k}-\eps^{e}_{k-p} - \hbar\Omega_{p,p_\perp})
           \right]\,,
\label{eq:definegammatildee}
\\
\gamma^{h}_{k,p} & = & 
           (-i) \sum_{p_\perp} |G_{p,p_\perp}|^2
           \left[
             (N^{\mrm{PH}}_{p,p_\perp}+1-f^{h}_{k-p}) 
             g(\eps^{h}_{k-p}-\eps^{h}_{k} + \hbar\Omega_{p,p_\perp})
           \right.
\nonumber\\
&&\qquad\qquad
+
           \left.
             (N^{\mrm{PH}}_{p,p_\perp}+f^{h}_{k-p}) 
             g(\eps^{h}_{k-p}-\eps^{h}_{k} - \hbar\Omega_{p,p_\perp})
           \right]\,.
\label{eq:definegammatildeh}
\end{eqnarray}
%%%%%%%%%%%%%%%%%%%%
%%%%%%%%%%%%%%%%%%%%
%
Similar equations for electron and hole correlations are also given in
App.~\ref{app:phonons}.

These six-point phonon scattering contributions lead to microscopic dephasing of
correlations and we find the conservation law 
%
%%%%%%%%%%%%%%%%%%%%
%%%%%%%%%%%%%%%%%%%%
\begin{equation}
\left[
\ddt \sum_{k,k',q} c^{q,k',k}_{\lambda} 
\right]_{H_{\mrm{P}}} = 0
\label{eq:corrcons}
\end{equation}
%%%%%%%%%%%%%%%%%%%%
%%%%%%%%%%%%%%%%%%%%
%
for the complex valued correlations $c^{q,k',k}_{\lambda}$ such that
the phonon scattering has a diffusive character.
This property allows formation as well as equilibration of correlations. 

The set of equations~(\ref{eq:cvcv})--(\ref{eq:SBEvv}), (\ref{eq:cccc}), and~(\ref{eq:vvvv}),
extended by the phonon contributions, Eqs.~(\ref{eq:ddtfephon2})--(\ref{eq:ddtcXphon6p}),
(\ref{eq:ddtcephon6p}) and~(\ref{eq:ddtchphon6p}), provides
a closed system of coupled differential equations which can be solved under
different initial conditions in order to study exciton formation for different carrier
densities and lattice temperatures.
%
%%%%%%%%%%%%%%%%%%%%%%%%%%%%%%%%%%%%%%%%%%%%%%%%%%%%%%%%%%%%%%%%%%%%%%%%%%%
%%%%%%%%%%%%%%%%%%%%%%%%%%%%%%%%%%%%%%%%%%%%%%%%%%%%%%%%%%%%%%%%%%%%%%%%%%%
%
%
%%%%%%%%%%%%%%%%%%%%%%%%%%%%%%%%%%%%%%
\section{Exciton Formation}
\label{sec:exform}
%%%%%%%%%%%%%%%%%%%%%%%%%%%%%%%%%%%%%%
%
Because of the high dimensionality of the summations in our hierarchy of
coupled equations, the currently available computer resources limit us to
evaluate the equations for a one-dimensional model system. However, we choose parameters
to be close to the standard GaAs parameters used for quantum wells. The effective width of our
quantum wire is determined such that the exciton binding energy lies roughly 11\,meV below the
unrenormalized bandgap which is also the case in typical 8\,nm quantum wells.
The corresponding 3D-exciton Bohr radius is $a_0=12.4\,$nm. To incorporate Coulomb
screening effects to the four-point terms, we use a statically screened Coulomb
potential obtained from the Lindhard formula.\cite{Haug:94} Microscopically, the
justification follows from the coupling to the six-point terms. For the carrier densities,
the coupling to the doublet level provides a microscopic description of scattering and
screening such that the unscreened matrix element is used in Eq.~(\ref{eq:SBEcc})
and~(\ref{eq:SBEvv}). In all calculations, we start from an incoherent electron-hole plasma
with vanishing correlated doublets.
%
%%%%%%%%%%%%%%%%%%%%%%%%%%%%%%%%%%%
\subsection{Direct Evidence of Exciton Formation}
\label{sec:formation}
%%%%%%%%%%%%%%%%%%%%%%%%%%%%%%%%%%%
%
The formation of any significant amount of bound excitons can be directly observed
via the electron-hole pair-correlation function\cite{Hoyer:02d,Zimmermann:88}
%
%%%%%%%%%%%%%%%%%%%%%%%%%
\be
 g^{eh}(r) \equiv
\frac{1}{n^2}
 \ev{
     \Psi^{\dagger}_{e}(r) \Psi^{\dagger}_{h}(0)
     \Psi_{h}(0) \Psi_{e}(r)
 },
\label{eq:paircorr}
\ee
%%%%%%%%%%%%%%%%%%%%%%%%%
%
which is normalized with respect to the single-particle carrier density $n$ and
measures the conditional probability to find an electron at position
$r$ while the hole is located at $r=0$. The pair-correlation function can be
separated into a singlet and a correlated doublet part
%
%%%%%%%%%%%%%%%%%%%%
\bea
g^{eh}(r)
& = &
\frac{1}{(n\mcal{L})^{2}}
\sum_{k,k',p,p'} e^{-i (k-k') r}
        \ev{c^{\dagger}_{k} c_{k'} v_{p} v^{\dagger}_{p'}}
=
g^{eh}_{\mrm{S}}(r) + \Delta g^{eh}(r,t),
\label{eq:paircorr2}
\\
g^{eh}_{\mrm{S}}(r) 
& = &
1 + \left|\frac{1}{n\mcal{L}}\sum_{k} e^{i k r} P_k\right|^2,
\label{eq:geh_S}
\\
\Delta g^{eh}(r)
& = &
\frac{1}{(n\mcal{L})^{2}}\sum_{k,k',q} e^{-i (k-k'-q) r} \;  c_{X}^{q,k',k},
\label{eq:Deltageh}
\eea
%%%%%%%%%%%%%%%%%%%%
%
where $P_{k} = \langle a^{\dagger}_{v,k} a_{c,k}\rangle$ is the coherent microscopic
interband transition amplitude which provides the dominant contribution to the
pair-correlation function after a coherent excitation with a classical field.\cite{Vu:00}
As soon as this polarization dephases, however, the
factorized contribution becomes constant and the only $r$-dependence is
provided by the correlated part $\Delta g^{eh}(r)$.
This quantity provides an intuitive measure of when and how fast exciton formation and
ionization takes place in incoherent configurations.

%
%%%%%%%%%%%%%%%%%%%%%
\begin{figure}[th]
\includegraphics[width=0.5\textwidth]{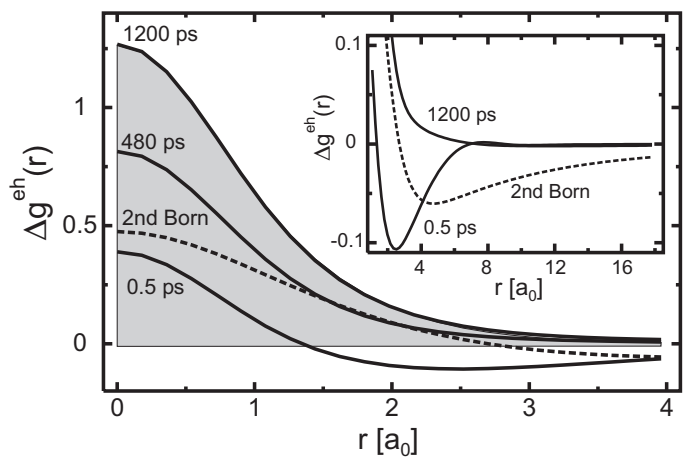}
\caption{Pair-correlation function $\Delta g^{\mrm{eh}}(r)/(n^{\mrm{e}} n^{\mrm{h}})$ 
normalized to the constant Hartree-Fock value for a lattice temperature of $T=10$\,K and
a carrier density of $n=2\times{}10^{4}$\,cm$^{-1}$ at different times.
For comparison, the wave function of the lowest bound exciton
is given as a shaded area. The dashed line is the corresponding result of a 
second-Born computation. The inset shows a magnification of the tails of the same curves.}
\label{fig:pair}
\end{figure}
%%%%%%%%%%%%%%%%%%%%%
%
Figure~\ref{fig:pair} shows computed examples of $\Delta g^{eh}(r)$
as a function of electron-hole distance~$r$
for a low carrier density of $n^{{e/h}}=2\times 10^{4}\,\mrm{cm}^{-1}$. At early
times after the completely uncorrelated initialization of
the system, the carriers react to the Coulomb attraction and
the probability of finding electrons and holes close to each other increases.
However, the correlated $\Delta g^{eh}$ has clearly
negative parts. At certain distances, the total pair correlation $g^{eh}(r)$ is
therefore below its factorized value. We interpret this behavior as a fast rearrangement
of the electron-hole plasma where the probability to find electrons in the close vicinity
of holes is increased at the expense of the probability at around 2 to 4 $a_0$. Since the
area under $\Delta g^{eh}$ provides a measure of how many excitons 
have formed\cite{Hoyer:02d}, the early time dynamics does not show a real formation
of excitons out of the plasma. Nevertheless, at 480\,ps, the pair correlation has assumed
the shape of an exciton relative-motion wave function (given as a shaded area) and grows almost
linearly in time.

For comparison, we also show the second Born result in Fig.~\ref{fig:pair} as a dashed line.
In this approximation, the Coulomb sums of Eqs.~(\ref{eq:cvcv}) and the triplet phonon scattering,
Eq.~(\ref{eq:ddtcXphon6p}), are not included such that the doublets can describe only
carrier-carrier scattering but not bound excitons. We see that the resulting pair-correlation
function has similar characteristics as the early time $\Delta g^{eh}$ in the full computation.
In particular, $\Delta g^{eh}$ drops to negative values before it approaches zero at large
distances. Since no exciton populations are included in the second-Born computation,
the resulting shape of the pair correlation has to be interpreted as a rearrangement
within the plasma.

%
%%%%%%%%%%%%%%%%%%%%%%%%%%%%%%%%%%%%%%
\subsection{Phonon Induced Energy Transfer in Exciton Formation}
\label{sec:energytransfer}
%%%%%%%%%%%%%%%%%%%%%%%%%%%%%%%%%%%%%%
%
Since a bound exciton at rest has a lower energy than a free electron-hole pair, we
study the energy transfer during the exciton formation in the many-body system.
For carriers alone, the system energy is
%
%%%%%%%%%%%%%%%%%%%%
\begin{eqnarray}
\ev{H_{\mrm{kin}}}+ \ev{H_{\mrm{C}}} &=& 
\sum_{\lambda,k} 
\eps^{\lambda}_{k} \ev{a^{\dagger}_{\lambda,k} a_{\lambda,k}}
+
\frac{1}{2}\sum_{\lambda,\lambda' \atop k,k',q\not=0} 
V_{q} 
\ev{a^{\dagger}_{\lambda,k} a^{\dagger}_{\lambda',k'} a_{\lambda',k'+q} a_{\lambda,k-q}}
\nonumber\\
& = &
\ev{H^{e}_{\mrm{kin}}} + \ev{H^{h}_{\mrm{kin}}}
+ \ev{H^{e}_{\mrm{C}}}_{\mrm{S}}
+ \ev{H^{h}_{\mrm{C}}}_{\mrm{S}}
\nonumber\\
&&\qquad\qquad
+ \Delta \ev{H^{e}_{\mrm{C}}} 
+ \Delta \ev{H^{h}_{\mrm{C}}}
+ \Delta \ev{H^{eh}_{\mrm{C}}}
\label{eq:energysum}
\end{eqnarray}
%%%%%%%%%%%%%%%%%%%%
%
with the kinetic energies
$\ev{H^{\lambda}_{\mrm{kin}}} = \sum_{k} \eps^{\lambda}_{k} f^{\lambda}_{k}$,
the incoherent singlet (factorized) parts to the Coulomb energy
$\ev{H^{\lambda}_{\mrm{C}}}_{\mrm{S}} =
-
\frac{1}{2}\sum_{k,k'\not=k} V_{k-k'} 
f^{\lambda}_{k'} f^{\lambda}_{k}$,
and the correlated contributions
$\Delta \ev{H^{\lambda}_{\mrm{C}}} 
=
\frac{1}{2}\sum_{k,k',q\not=0} V_{q} c_{\lambda}^{q,k',k}$
and
$\Delta \ev{H^{eh}_{\mrm{C}}}
=
- \sum_{k,k',q\not=0} V_{q} c_{X}^{k-q-k',k',k}$.
Using Eqs.~(\ref{eq:cvcv})--(\ref{eq:SBEvv}) we find
%
%%%%%%%%%%%%%%%%%%%%
%%%%%%%%%%%%%%%%%%%%
\be
\ddt \left[
\langle H_{\mrm{kin}} \rangle
+
\langle H_{\mrm{C}} \rangle
\right]_{H_{\mrm{kin}}+H_{\mrm{C}}}
=0
\ee
%%%%%%%%%%%%%%%%%%%%
%%%%%%%%%%%%%%%%%%%%
%
which shows that the energy is conserved within the isolated carrier system.
As a result, exciton formation is not efficient for a plain carrier system. When
also phonons are included, we find
%
%%%%%%%%%%%%%%%%%%%%
%%%%%%%%%%%%%%%%%%%%
\be
\ddt \langle H_{\mrm{tot}} \rangle = 
\ddt 
\left[
\langle H_{\mrm{kin}} \rangle
+
\langle H_{\mrm{C}} \rangle
+
\langle H_{\mrm{phon}} \rangle
+
\langle H_{\mrm{P}} \rangle
\right]
=0\,,
\label{eq:Etotcons}
\ee
%%%%%%%%%%%%%%%%%%%%
%%%%%%%%%%%%%%%%%%%%
%
i.e., the truncation via the cluster expansion scheme fully conserves
the total energy of the system. When the phonons are included, the carrier
system can cool down because part of the energy gained via exciton formation is transferred
to lattice vibrations. When the phonons are treated as a reservoir, the phonon bath
acts as a sink and the energy flux directed to the phonon system is absorbed by the
reservoir. However, when we replace the microscopic phonon scattering,
Eq.~(\ref{eq:ddtcXphon6p}), by a simplified constant dephasing rate $\gamma$, we find
%
%%%%%%%%%%%%%%%%%
\be
\ddt \ev{H}_{\mrm{tot}} = -\gamma \,\Delta\ev{H_{\mrm{C}}} > 0
\ee
%%%%%%%%%%%%%%%%%
%
which implies an unphysical heating of the system. Therefore, 
systematic exciton formation
studies require a microscopic description of phonon scattering.

%
%%%%%%%%%%%%%%%%%%%%%
\begin{figure}[bt]
\includegraphics[width=0.3\textwidth]{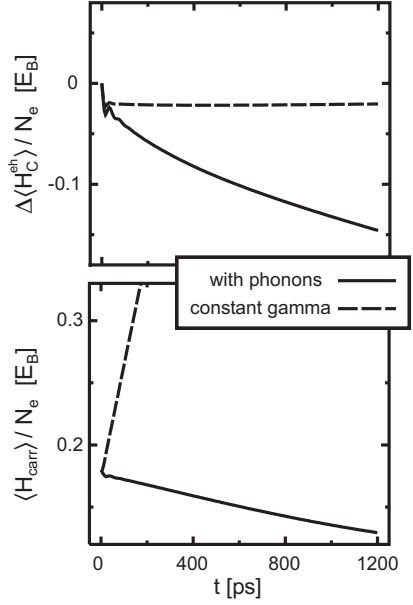}
\caption{Attractive Coulomb energy (a) and total energy (b) per particle comparing a computation with
constant dephasing approximation (dashed line) with the full result including microscopic
phonon scattering (solid line) for the same parameters as in Fig.~\ref{fig:pair}.
Energies are given in multiples of the 3D exciton binding energy $E_{\mrm{B}}=4.2$\,meV.}
\label{fig:energy}
\end{figure}
%%%%%%%%%%%%%%%%%%%%%
%
The dynamic evolution of the carrier energy is shown in Fig.~\ref{fig:energy} by using 
either constant dephasing or microscopic phonon scattering. With constant $\gamma$,
the attractive Coulomb energy very quickly reaches its steady state value such that
significant pair formation is not observed. Even a small constant $\gamma$ of
50\,$\mu$eV leads to a linear heating of the carrier distributions, i.e.,
to an increase of the total energy. With the microscopic phonon scattering and a
lattice temperature of 10\,K, however, true formation of
excitons is possible as indicated by the continuous decrease of the Coulomb energy.
Simultaneously, the total energy of the carrier system decreases since part of the
energy is lost to the phonon bath.
The slope of decrease of the attractive Coulomb energy can in 
principle be used as a measure of how fast the exciton formation takes place.

%
%%%%%%%%%%%%%%%%%%%%%
\begin{figure}[!ht]
\includegraphics[width=0.5\textwidth]{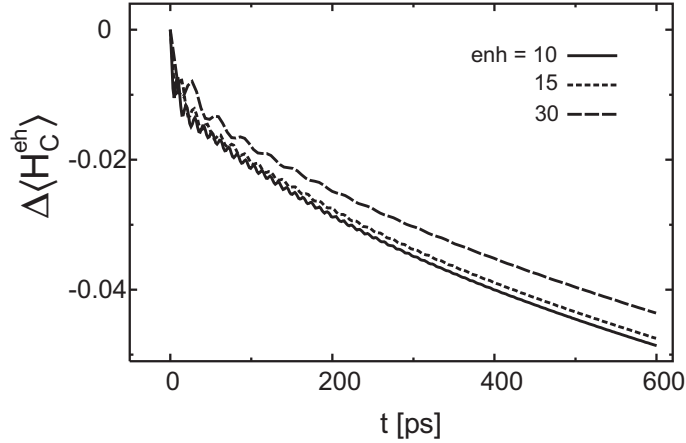}
\caption{Correlated Coulomb energy $\Delta\ev{H^{\mrm{eh}}_{\mrm{C}}}$ per particle for three different 
computations with different enhancement factors. For each curve, the time axis has been
rescaled with the  respective enhancement factor. The carrier density is $n=1\times{}10^{4}$\,cm$^{-1}$
and the lattice temperature is $T=10$\,K.}
\label{fig:enhance}
\end{figure}
%%%%%%%%%%%%%%%%%%%%%
%
To speed up the numerics, we have computed
Fig.~\ref{fig:energy} by following the dynamics upto 40\,ps with a 
phonon matrix-element enhanced by a factor of 30; this corresponds to 
$30\times 40\,\mrm{ps}=1.2\,\mrm{ns}$ of formation dynamics. To verify
the validity of this analysis we show in
Fig.~\ref{fig:enhance} a comparison for
three enhancement factors of 10, 15, and 30.  After rescaling the time axis with the respective
factor, the behavior of the correlated Coulomb energy is indeed independent of the enhancement
factor. In particular, the rate of change is very similar for all three cases which 
justifies the use of the artificial enhancement factors in the numerical investigations.

%
%%%%%%%%%%%%%%%%%%%%%%%%%%%%%%%%%%%%%%
\subsection{Formation of Specific Excitons}
\label{sec:adiabatic}
%%%%%%%%%%%%%%%%%%%%%%%%%%%%%%%%%%%%%%
%

In the equation for excitonic correlations Eq.~(\ref{eq:cvcv}), two of the
six Coulomb sums are usually dominant; the terms multiplied by the phase
space filling factor $1-f^{e}-f^{h}$ provide a large contribution
for all densities. Therefore, the restriction to these two dominant
sums is often a very good approximation to the full result.
The corresponding ``main sum approximation'' of Eq.~(\ref{eq:cvcv}) is
%
%%%%%%%%%%%%%%%%%%%%%%%%%%%%%%%%%%%%%%%%%%%%%%%%%%%%%%%%%%%%%%%%%%%%%%%%%%%
\begin{eqnarray}
\lefteqn{%
i\hbar\ddt c_{X}^{q,k',k} = 
                (
        \tilde{\eps}_{k'+q}^{e}-\tilde{\eps}_{k-q}^{h}
                -
        \tilde{\eps}_{k}^{e}+\tilde{\eps}_{k'}^{h}
                )
        c_{X}^{q,k',k}}
\nonumber\\
 && +{} V_{k-k'-q}
                \left[
        (1-f_{k}^{e})(1-f_{k-q}^{h})f_{k'+q}^{e}f_{k'}^{h}
                -
        (1-f_{k'+q}^{e})(1-f_{k'}^{h})f_{k}^{e}f_{k-q}^{h}
                \right]
\nonumber\\
 && +   [1-f_{k}^{e}-f_{k-q}^{h}] \sum _{l}V_{l-k}c_{X}^{q,k',l}
                -
        [1-f_{k'+q}^{e}-f_{k'}^{h}]\sum _{l}V_{l-k'}c_{X}^{q,l,k}\,.
\label{eq:mainsum}
\end{eqnarray}
%%%%%%%%%%%%%%%%%%%%%%%%%%%%%%%%%%%%%%%%%%%%%%%%%%%%%%%%%%%%%%%%%%%%%%%%%%%
%
With this restriction, the excitonic correlations are not coupled to 
carrier-carrier correlations anymore and we obtain a closed subsystem of
equations for each center-of-mass momentum $q$.

Since the carrier distributions typically vary slowly, we try to find an
adiabatic solution for Eq.~(\ref{eq:mainsum}). To do so, we use a generalized
exciton basis, which is defined by the equations
%
%%%%%%%%%%%%%%%
\bea
\tilde\eps_{k,q} \phi^{r}_{\nu,q}(k)
        -
(1-f_{k+q^{e}}^{e}-f_{k-q^{h}}^{h})
\sum_{k'} V_{k'-k}\phi^{r}_{\nu,q}(k') 
        & = &
E_{\nu,q}\phi^{r}_{\nu,q}(k)\,,
\label{eq:wannier}
\\
\left(
\phi^{l}_{\nu,q}(k)
\right)^*
\tilde\eps_{k,q}
        -
\sum_{k'}
\left(
\phi^{l}_{\nu,q}(k')
\right)^*
V_{k'-k} (1-f_{k'+q^{e}}^{e}-f_{k'-q^{h}}^{h})
        & = &
\left(
\phi^{l}_{\nu,q}(k)
\right)^*
E_{\nu,q}\,,
\label{eq:wannierleft}
\eea
%%%%%%%%%%%%%%%
%
where we have defined 
$\tilde\eps_{k,q} = \tilde{\eps}_{k+q^{e}}^{e} +\tilde{\eps}_{k-q^{h}}^{h}$
and $q^{e(h)} = q\,m_{{e(h)}}/(m_{e} + m_{h})$.
The main difference to the usual Wannier basis is that for non-zero densities
the phase-space filling factor and a screened Coulomb potential enter the
effective Hamiltonian. Therefore, the eigenvalue problem Eq.~(\ref{eq:wannier})
is not Hermitian and one obtains left and right handed eigenfunctions $\phi^{l(r)}_{\nu,q}$.
This exciton basis fulfills a generalized orthogonality and completeness relation
%
%%%%%%%%%%%%%%%%%%%
\bea
\sum_{k} \left(\phi^{l}_{\nu,q}(k)\right)^* \phi^{r}_{\nu',q}(k)
&=&
\delta_{\nu,\nu'},
\label{eq:xortho}
\\
\sum_{\nu} \left(\phi^{l}_{\nu,q}(k)\right)^* \phi^{r}_{\nu,q}(k')
&=&
\delta_{k,k'}.
\label{eq:xcomplete}
\eea
%%%%%%%%%%%%%%%%%%
%
The center-of-mass momentum $q$ enters Eq.~(\ref{eq:wannier}) via the phase-space
filling factor such that the relative motion of an exciton depends also on its
center-of-mass momentum.

By means of the generalized Wannier functions, one can introduce exciton annihilation operators
%
%%%%%%%%%%%%%%%%%%%%%%%
\bea
X_{\nu,q}
& = &
\sum_{k} \left(\phi^l_{\nu,q}(k)\right)^* v^{\dagger}_{k-q^h} c_{k+q^e},
\label{eq:exciton}
\\
v^{\dagger}_{k-q^h} c_{k+q^e}
& = &
\sum_{\nu} \phi^r_{\nu,q}(k) \; X_{\nu,q}
\label{eq:excinv}
\eea
%%%%%%%%%%%%%%%%%%%%%%%
%
and the Hermitian conjugate creation operators.
Equations~(\ref{eq:exciton}) and~(\ref{eq:excinv}) can directly be applied
to the correlations in order to transform them according to
%
%%%%%%%%%%%%%%%%%%%%%%%%%%%%%%%%%%%%%%%%%%%%%%%%%%%%%%%%%%%%%%%%%%%%%%%%%%%
\bea
c_{X}^{q,k'-q^{h},k+q^{e}}
        & = & 
\sum_{\nu,\nu'} \left(\phi^{r}_{\nu,q}(k)\right)^* \phi^{r}_{\nu',q}(k')\,
        \Delta\ev{X^{\dagger}_{\nu,q} X_{\nu',q}},
\label{eq:cXinX}
\\
\Delta\ev{X^{\dagger}_{\nu,q} X_{\nu',q}}
        & = &
\sum_{k,k'} \phi^{l}_{\nu,q}(k) \left(\phi^{l}_{\nu',q}(k')\right)^*\,
c_{X}^{q,k'-q^{h},k+q^{e}}.
\label{eq:defXX}
\eea
%%%%%%%%%%%%%%%%%%%%%%%%%%%%%%%%%%%%%%%%%%%%%%%%%%%%%%%%%%%%%%%%%%%%%%%%%%%
%
This expansion cannot only be used to solve Eq.~(\ref{eq:mainsum}) but also to
compute expectation values in the exciton basis from our numerical computations
performed in the $k$-basis.

%
%%%%%%%%%%%%%%%%%%%%%
\begin{figure}[!ht]
\includegraphics[width=0.45\textwidth]{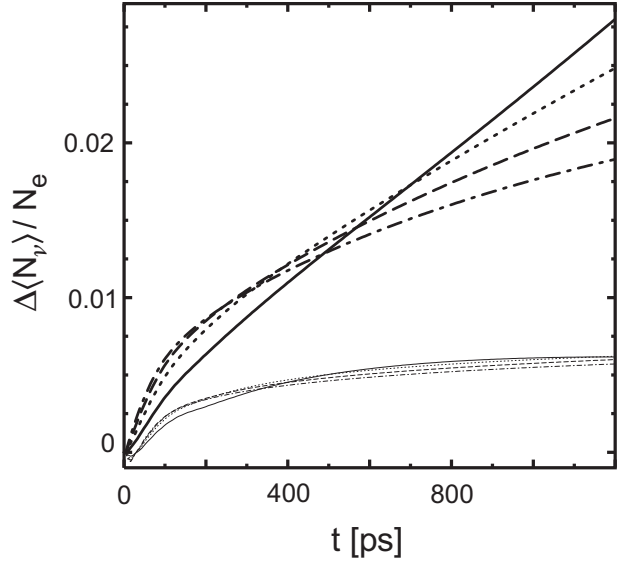}
\caption{Formation dynamics of $q$-integrated population correlations, Eq.~(\ref{eq:Nnutot}),
out of an electron-hole plasma with $n=2\times 10^{4}\,\mrm{cm}^{-1}$ at $T=60$\,K.
The lattice temperature was 10\,K (solid line),
20\,K (dotted line), 30\,K (dashed line), and 40\,K (dash-dotted line), respectively. Thick
lines indicate $\Delta\ev{N_{\mrm{1s}}}$ whereas thin lines refer to $\Delta\ev{N_{\mrm{2p}}}$.}
\label{fig:T60}
\end{figure}
%%%%%%%%%%%%%%%%%%%%%
%
Using Eq.~(\ref{eq:defXX}),
we can study the formation of specific excitons as shown in Fig.~\ref{fig:T60}. 
There, we have computed the dynamics for different lattice temperatures, always starting
with initial electron and hole distributions at a  temperature of $T=60$\,K. With the
help of Eq.~(\ref{eq:defXX}), we compute the $q$-integrated exciton population correlations
%
%%%%%%%%%%%%%%%%%
\be
\Delta\ev{N_\nu} = \sum_{q} \Delta\ev{N_{\nu,q}} = \sum_{q} \Delta\ev{X^{\dagger}_{\nu,q}X_{\nu,q}}
\label{eq:Nnutot}
\ee
%%%%%%%%%%%%%%%%%
%
normalized to the total number of electrons in the system. These populations as a function
of time are shown for the two lowest bound excitons denoted 1s and 2p, respectively%
\footnote{In a one-dimensional model system, all exciton wave functions have a defined
parity and are either even or odd functions. In analogy to the common nomenclature, we
label the bound states 1s, 2p, 3s, and so forth.}. Figure~\ref{fig:T60} clearly
demonstrates how the formation rate of 1s excitons quickly drops for elevated lattice
temperatures. For temperatures above 40\,K, we do not expect any significant formation.

In order to continue our analytical derivation, we note that in the incoherent regime
$f^{e}$ and  $f^{h}$ typically change slowly such that the exciton wave functions 
$\phi_{\nu,q}$ can be assumed to be quasi stationary. With the help of this adiabatic
approximation, Eq.~(\ref{eq:mainsum}) can be written as
%
%%%%%%%%%%%%%%%%%%%%%%%%%%%%%%%%%%%%%%%%%%%%%%%%%%%%%%%%%%%%%%%%%%%%%%%%%%%
\bea
i\hbar\ddt
\Delta\ev{X^{\dagger}_{\nu,q} X_{\nu',q}}
        & = &
(E_{\nu'}-E_{\nu}) \Delta\ev{X^{\dagger}_{\nu,q} X_{\nu',q}}
\nonumber\\
&& \qquad\quad
+{} \sum_{k,k'} \phi^{l}_{\nu,q}(k) \left(\phi^{l}_{\nu',q}(k')\right)^* 
                                        S^{q,k',k}
\label{eq:cvcvX}
\eea
%%%%%%%%%%%%%%%%%%%%%%%%%%%%%%%%%%%%%%%%%%%%%%%%%%%%%%%%%%%%%%%%%%%%%%%%%%%
%
with the factorized (singlet) source term
%
%%%%%%%%%%%%%%%%%%%%%%%%%%%%%%%%%%%%%%%%%%%%%%%%%%%%%%%%%%%%%%%%%%%%%%%%%%%
\bea
S^{q,k',k}
& = &
V_{k-k'}        \left[
        (1-f^{e}_{k+q^{e}}-f^{h}_{k-q^{h}})
        f^{e}_{k'+q^{e}} f^{h}_{k'-q^{h}}
                \right.
\nonumber\\
&&\qquad\qquad\qquad
                -
                \left.
        (1-f^{e}_{k'+q^{e}}-f^{h}_{k'-q^{h}})
        f^{e}_{k+q^{e}} f^{h}_{k-q^{h}}
                \right]\,.
\label{eq:source}
\eea
%%%%%%%%%%%%%%%%%%%%%%%%%%%%%%%%%%%%%%%%%%%%%%%%%%%%%%%%%%%%%%%%%%%%%%%%%%%
%
The properties of the left-handed exciton basis, Eq.~(\ref{eq:wannierleft}),
can be used to simplify the last term in Eq.~(\ref{eq:cvcvX}) according to
%
%%%%%%%%%%%%%%%%%%%%%%%%
\bea
%\lefteqn{%
\sum_{k,k'} \phi^{l}_{\nu,q}(k) \left(\phi^{l}_{\nu',q}(k')\right)^* 
                                        S^{q,k',k}
        & = &
(E_{\nu'} - E_{\nu}) \ev{X^{\dagger}_{\nu,q} X_{\nu',q}}_{\mrm{S}}\,,
\eea
%%%%%%%%%%%%%%%%%%%%%%%%
%
where the factorized plasma part of the two-particle correlation is given by
%
%%%%%%%%%%%%%%%%%%%%%%%%
\be
\ev{X^{\dagger}_{\nu,q} X_{\nu',q}}_{\mrm{S}} =
\sum_{k} \phi^{l}_{\nu,q}(k)
                        \left(\phi^{l}_{\nu',q}(k)\right)^* 
        f^{e}_{k+q^{e}} f^{h}_{k-q^{h}}\,.
\label{eq:XX2p}
\ee
%%%%%%%%%%%%%%%%%%%%%%%%
%       
In the exciton basis, the full equation is therefore
%
%%%%%%%%%%%%%%%%%%%%%%%%
\bea
i\hbar\ddt
\Delta\ev{X^{\dagger}_{\nu,q} X_{\nu',q}} & = & 
        (E_{\nu'}-E_{\nu}) \Delta\ev{X^{\dagger}_{\nu,q} X_{\nu',q}}
        +
        (E_{\nu'}-E_{\nu}) \ev{X^{\dagger}_{\nu,q} X_{\nu',q}}_{\mrm{S}}
\nonumber\\
&&
+i\hbar
\left[
\ddt
\Delta\ev{X^{\dagger}_{\nu,q} X_{\nu',q}}
\right]_{\mrm{T-scatt}}
\label{eq:adiabatic}
\eea
%%%%%%%%%%%%%%%%%%%%%%%%
%
where we have included the scattering contributions from triplet correlations.

The simple form of Eq.~(\ref{eq:adiabatic}) reveals how the exciton formation proceeds.
If we start the calculations assuming initially vanishing correlations,
they start to build up because the source term $\ev{X^{\dagger}_{\nu,q} X_{\nu',q}}_{\mrm{S}}$
is non-vanishing as soon as carriers are present in the system. However, that source term
does not create diagonal correlations
such that excitonic populations $\Delta \ev{N_{\nu,q}}$ stay zero without the
six-point scattering contribution. Hence, the formation of exciton populations in the 
incoherent regime has only one possible Coulombic channel where off-diagonal transition
correlations $\Delta\ev{X^{\dagger}_{\nu,q} X_{\nu'\not=\nu,q}}$ are created first.
Only after off-diagonal transitions have been created, populations can be formed via the triplet
scattering. Therefore, it is crucial to include both excitonic transition and population
correlations in the formation analysis, as is
done automatically in the full $k$-basis treatment used in all our computations.
If one uses excitonic models which are restricted to the diagonal
populations, the formation channel via the transition correlations is completely
omitted. 

%
%%%%%%%%%%%%%%%%%%%%%
\begin{figure}[!ht]
\includegraphics[width=0.45\textwidth]{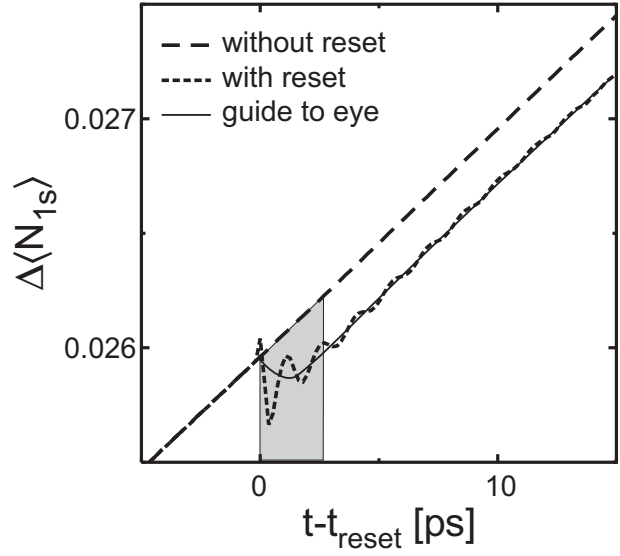}
\caption{Total exciton population correlation $\sum_{q}\Delta\ev{N_{\mrm{1s},q}}$
for two computations with (dotted) and without (dashed) reset of off-diagonal transition
correlations for $n=2\times 10^{4} \,\mrm{cm}^{-1}$ and $T=10$\,K.
The thin solid line is given as guide to the eye.}
\label{fig:switchoff}
\end{figure}
%%%%%%%%%%%%%%%%%%%%%
%
The efficiency and the character of the indirect exciton formation channel are analyzed in
Fig.~\ref{fig:switchoff} which investigates the build-up of the integrated exciton
population correlation $\Delta\ev{N_{\mrm{1s}}}$. Here, we compare a run where all
off-diagonal transition correlations $\Delta\ev{X^{\dagger}_{\nu} X_{\nu'\not=\nu}}$
are set to zero during the computation with a reference run without reset. For the reset,
all correlations are transformed from the $k$-basis into the exciton basis via
Eq.~(\ref{eq:defXX}) and back to the $k$-basis with the help of Eq.~(\ref{eq:cXinX}). In the
back transformation, we omit all but the diagonal correlations.
No phonon-enhancement factor was used in this computation
because we want to investigate Coulomb and phonon dynamics for equally short time scales.
Exactly at the moment where the off-diagonal correlations are reset, the formation dynamics
stops and no more exciton correlations can build up. On the contrary, the amount of 1s
populations is even slighly decreased 
since the phonon scattering tends to transfer some of the
diagonal populations back to off-diagonal transition correlations. Already around 3\,ps
after the reset, however, the off-diagonal transition correlations are fully recovered. 
After some transient dynamics, the formation continues with the previous formation
rate. Also in other quantities, as for example in the
correlated Coulomb energy, this fast recovery of the transition correlations is confirmed. 
This behavior indicates that off-diagonal transition correlations build up
extremely quickly compared to the formation rate of diagonal populations. It is these
off-diagonal transition correlations which cause the plasma like shape of the early-time
pair-correlation function.

The importance of the off-diagonal exciton transitions can also be seen in the
equation of motion of the electron density.
By expressing the right hand side of Eq.~(\ref{eq:SBEcc}) with the help of
Eq.~(\ref{eq:cXinX}), we obtain
%
%%%%%%%%%%%%%%%%%%%%%%%%
\be
\ddt f^{e}_{k}\Bigr|_{\mrm{X-corrs}} = \sum_{\nu,\nu',q} (E_{\nu',q}-E_{\nu,q})
        (\phi^{l}_{\nu',q}(k-q^e))^* \phi^{r}_{\nu,q}(k-q^e)
        \Delta\ev{X^{\dagger}_{\nu,q}X_{\nu',q}}\,.
\label{eq:ddtfeX}
\ee
%%%%%%%%%%%%%%%%%%%%%%%%
%
Thus, diagonal exciton populations do not change the carrier densities. Excitonic
transitions are required in order to describe the full dynamics and the correct heating
of the carrier plasma due to exciton formation.

%
%%%%%%%%%%%%%%%%%%%%%%%%%%%%%%%%%%%%%%
\subsection{Transition Correlations}
\label{sec:offdiagonal}
%%%%%%%%%%%%%%%%%%%%%%%%%%%%%%%%%%%%%%
%
In order to understand the general nature of the off-diagonal transition correlations,
it is sufficient to investigate Eq.~(\ref{eq:adiabatic}) where the six-point scattering
contributions are approximated by a constant dephasing rate, i.e., 
%
%%%%%%%%%%%%%%%
\be
\left[
\hbar\ddt \ev{X^{\dagger}_{\nu,q} X_{\nu',q}}
\right]_{\mrm{T-scatt}}
\approx
-\gamma\, \ev{X^{\dagger}_{\nu,q} X_{\nu',q}}\,.
\ee
%%%%%%%%%%%%%%%
%
Since this term leads to a simple decay, Eq.~(\ref{eq:adiabatic}) has a steady-state
solution
%
%%%%%%%%%%%%%%%%%%%%%%%%
\be
\Delta\ev{X^{\dagger}_{\nu,q} X_{\nu',q}}\Bigr|_{\mrm{steady-state}}
        =
- \frac{E_{\nu'} - E_{\nu}}{E_{\nu'} - E_{\nu}-i\gamma} 
                 \ev{X^{\dagger}_{\nu,q} X_{\nu',q}}_{\mrm{S}}\,.
\label{eq:steadystate}
\ee
%%%%%%%%%%%%%%%%%%%%%%%%
%
For sufficiently small $\gamma$, we find
%
%%%%%%%%%%%%%%%%%%%%%%%%
\be
\Delta\ev{X^{\dagger}_{\nu,q} X_{\nu',q}}
        =
(\delta_{\nu,\nu'}-1)  \ev{X^{\dagger}_{\nu,q} X_{\nu',q}}_{\mrm{S}}\,.
\label{eq:steadystate2}
\ee
%%%%%%%%%%%%%%%%%%%%%%%%
%
For $\nu{\not=}\nu'$, the off-diagonal correlations
$\Delta\ev{X^{\dagger}_{\nu,q} X_{\nu',q}}=-\ev{X^{\dagger}_{\nu,q} X_{\nu',q}}_{\mrm{S}}$
fully cancel with the singlet (two-point) contribution when the total
$\ev{X^{\dagger}_{\nu,q} X_{\nu',q}}$ is evaluated. The
diagonal part $\ev{X^{\dagger}_{\nu,q} X_{\nu,q}}$, however, is determined
by the singlet contribution alone. In other words, we find
%
%%%%%%%%%%%%%%%%%%%%%%%%
\be
\ev{X^{\dagger}_{\nu,q} X_{\nu',q}} \approx \delta_{\nu,\nu'} 
        \ev{X^{\dagger}_{\nu,q} X_{\nu,q}}_{\mrm{S}} =
        \delta_{\nu,\nu'} 
        \sum_{k}  |\phi^{l}_{\nu,q}(k)|^{2}
        f^{e}_{k+q^{e}} f^{h}_{k-q^{h}}\,,
\label{eq:steadystate3}
\ee
%%%%%%%%%%%%%%%%%%%%%%%%
%
where the solution is expressed with the help of Eq.~(\ref{eq:XX2p}).

%
%%%%%%%%%%%%%%%%%%%%%
\begin{figure}[!ht]
\includegraphics[width=0.45\textwidth]{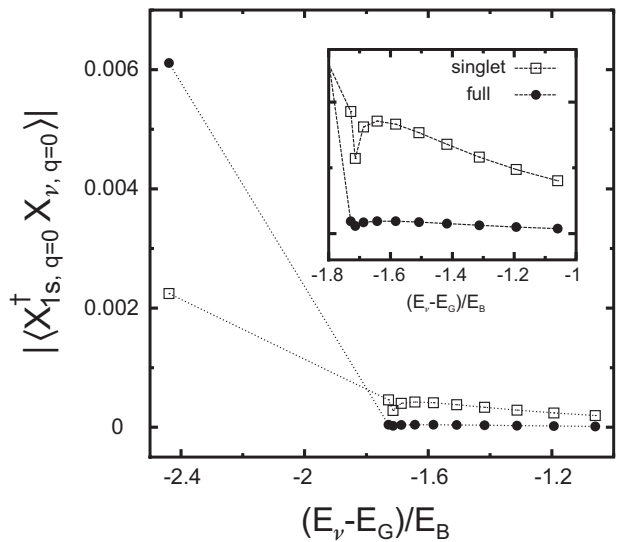}
\caption{Exciton correlations $\ev{X^{\dagger}_{1} X_{\nu}}$ as function of energy $E_{\nu}$
after 720\,ps of computation for the same parameters as in Fig.~\ref{fig:pair}.
The inset magnifies the region for energies higher
than $E_{\mrm{G}} - 1.8 E_{\mrm{B}}$. While off-diagonal correlations are largely reduced
in the full computation, the diagonal exciton population increases in time due to
genuine exciton formation.}
\label{fig:offdiag}
\end{figure}
%%%%%%%%%%%%%%%%%%%%%
%
Clearly, the off-diagonal transition correlations build up
to compensate the off-diagonal two-point contributions whereas diagonal
populations $\ev{X^{\dagger}_{\nu,q} X_{\nu,q}}$ have a non-vanishing and observable 
fermionic plasma contribution $\ev{X^{\dagger}_{\nu,q} X_{\nu',q}}_{\mrm{S}}$.
These observations are valid also for the more general case where the six-point phonon scattering
is included microscopically. Figure~\ref{fig:offdiag} shows the absolute value of the total
exciton correlations $|\ev{X^{\dagger}_{\mrm{1s},q=0} X_{\nu,q=0}}|$ as a function of
the exciton energy $E_{\nu}$ for such a general calculation with microscopic phonon
scattering. We compare the results of the full calculation with those obtained
if we truncate the equations at the singlet level where no excitonic correlations are
included.

The marker at the lowest energy gives the diagonal 1s exciton population
whereas the other markers indicate the magnitude of the respective off-diagonal transitions.
We see, that a calculation at the singlet level
predicts that the contributions of the off-diagonal transitions are
quite appreciable
(open squares) whereas the full calculation 
yields results that are more than an order
of magnitude smaller (full circles). Hence, we find a high degree of
cancellation between the singlet and the higher-order results
as predicted by Eq.~(\ref{eq:steadystate3}). Due to the very fast build-up
of off-diagonal correlations compared to the slow phonon scattering, the steady-state
result Eq.~(\ref{eq:steadystate2}) for off-diagonal exciton transitions is still a good
approximation for the full calculation. The main change to the simplified analysis leading
to Eq.~(\ref{eq:steadystate3}) is that under true formation conditions the diagonal
exciton populations can grow and exceed their singlet contribution. Due to the formation
dynamics, this diagonal population may increase in time and the difference to the plasma level
determines the amount of true populations. This difference can also be seen in Fig.~\ref{fig:offdiag}.

Even though the final correlations $\ev{X^{\dagger}_{\nu,q} X_{\nu',q}}$ are to a
very good approximation diagonal in the exciton basis, the distinction between
factorized and correlated part is very useful because the consistent
theory shows that: i) only the truly correlated part of the off-diagonal exciton
correlations leads to a significant contribution to the formation of excitons out of
an incoherent plasma, ii) only the off-diagonal correlations influence the time
evolution of the carrier distributions, and iii) even under good formation conditions,
the correlated contribution to the exciton population
is still of the same order of magnitude as its factorized counterpart. Both contributions
certainly influence experiments and must be carefully distinguished.
For example, the excitonic photoluminescence does not require the presence of
true excitonic populations $\Delta\ev{X^{\dagger}_{\nu,q} X_{\nu,q}}$.\cite{Kira:98}
%
%%%%%%%%%%%%%%%%%%%%%%%%%%%%%%%%%%%%%%%%%%%%%%%%%%%%%%%%%%%%%%%%%%%%%%%%%%%
%%%%%%%%%%%%%%%%%%%%%%%%%%%%%%%%%%%%%%%%%%%%%%%%%%%%%%%%%%%%%%%%%%%%%%%%%%%
%
%%%%%%%%%%%%%%%%%%%%%%%%%%%%%%%%%%%%%%%%%%%%%%%%%
%%%%%%%%%%%%%%%%%%%%%%%%%%%%%%%%%%%%%%%%%%%%%%%%%
\section{Summary}
\label{sec:summary}
%%%%%%%%%%%%%%%%%%%%%%%%%%%%%%%%%%%%%%%%%%%%%%%%%
%%%%%%%%%%%%%%%%%%%%%%%%%%%%%%%%%%%%%%%%%%%%%%%%%
%

We have presented a general many-body formalism to describe the dynamics of
charge carriers and two-particle correlations within semiconductor
heterostructures. The Coulomb interaction between electrons and holes and the
coupling of carriers and correlations to a phonon reservoir has been described 
microscopically. We thus are able to compute exciton formation times and
exciton-exciton correlations without neglecting the underlying fermionic properties.
We have shown that the formation is an intricate process where Coulomb correlations 
rapidly build up on a picosecond time scale while phonon dynamics leads to
true exciton formation on a slow nanosecond time scale.
An adiabatic approximative solution has been obtained which provides
an intuitive interpretation of the microscopic results.

%
%%%%%%%%%%%%%%%%%%%%%
\begin{figure}[!ht]
\includegraphics[width=0.5\textwidth]{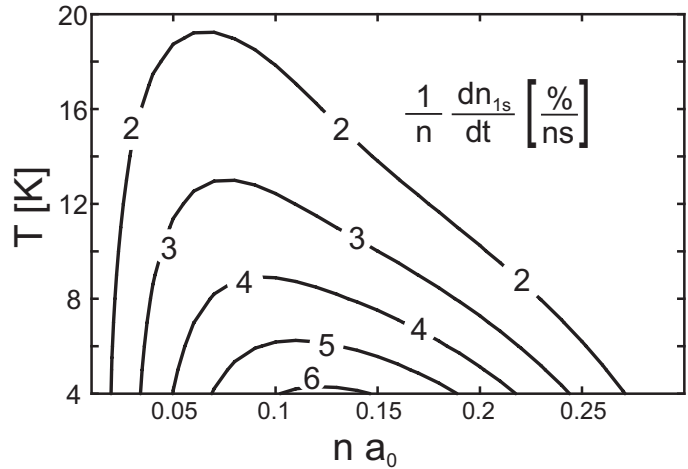}
\caption{Contour plot of exciton formation rates for an initial carrier temperature
of 60\,K and varying lattice temperatures and carrier densities. The exciton formation
rates are taken at 1.2\,ns of evolution and normalized by the carrier density. The contour
lines thus determine which percentage of charge carriers are bound to 1s-excitons
per nanosecond.}
\label{fig:contour}
\end{figure}
%%%%%%%%%%%%%%%%%%%%%
%
The basic results are summarized in Fig.~\ref{fig:contour}. There, we show the exciton formation
rate 1.2\,ns after initialization of the computations. At this time, the formation proceeds
almost linearly in time, c.f.\ Fig.~\ref{fig:T60}, and the formation rate gives a
good measure of the efficiency of the exciton formation. The contour lines show how many
percent of electron-hole pairs are bound to 1s-excitons per nanosecond.
Exciton formation is only efficient 
for cold lattice temperatures below 30\,K and rather low densities. For too low densities,
however, the rate of formation drops simply because it becomes increasingly improbable that
electrons and holes find scattering partners necessary for the formation process.

Generally, we can interpret the
formation of excitons as a build-up of correlations which
does not change the number of electrons and holes in the system. 
An interpretation of excitons as atom-like entities only makes sense under very restricted
conditions of a dilute gas where a small density of excitons interacts only very weakly
with the remaining carriers. Under those conditions, our theory gives reasonable exciton
numbers. Basically, the
total number of excitons formed is controlled by the lifetime of the electron-hole
excitations. In many realistic systems, this is in the order of ns because of
non-radiative recombination at imperfections and/or impurities.
%%%%%%%%%%%%%%%%%%%%%%%%%%%%%%%%%%%%%%%%%%%%%%%%%%%%%
%
%%%%%%%%%%%%%%%%%%%%%%%%%%%%%%%%%%%%%%%%%%%%%%%%%%%%%
\begin{acknowledgments}
This work was supported by the Deutsche Forschungsgemeinschaft, by the 
Max-Planck research prize of the Humboldt Foundation and the Max-Planck Society,  
and by the Forschungszentrum J\"{u}lich with a CPU-time grant.
M.K.~acknoledges funding from the Swedish Natural Science Research Council
(NFR) and the G\"{o}ran Gustafssons Stiftelse and thanks the Center for
Parallel Computers (PDC) to make their computer resources available
for this project. 
\end{acknowledgments}
%%%%%%%%%%%%%%%%%%%%%%%%%%%%%%%%%%%%%%%%%%%%%%%%%%%%%
%
\appendix
%
%%%%%%%%%%%%%%%%%%%%%%%%%%%%%%%%%%%%%%%%%%%%%%%%%%%%%
%
%%%%%%%%%%%%%%%%%%%%%%%%%%%%%%%%%%%%%%%%%%%%%%%%%%%%%%%%%%%%%%%%%%%%%%%%%%%%%
\section{Equations for electron and hole correlations}
\label{app:carrcorr}
%%%%%%%%%%%%%%%%%%%%%%%%%%%%%%%%%%%%%%%%%%%%%%%%%%%%%%%%%%%%%%%%%%%%%%%%%%%%%
%
In Section \ref{sec:eomHC} we have only presented the equation for the excitonic
correlations $c_{X}^{q,k',k}$. The remaining equations for electron-electron and hole-hole
correlations are 
%%%%%%%%%%%%%%%%%%%%%%%
%%%%%%%%%%%%%%%%%%%%%%%
\begin{eqnarray}
\lefteqn{
  \left[
  i \hbar\ddt
  c_{e}^{q,k',k} 
  \right]_{H_{\mrm{kin}}+H_{\mrm{C}}}
=
\left(
    \tilde{\epsilon}^{e}_{k-q}
    +
    \tilde{\epsilon}^{e}_{k'+q} 
    - 
    \tilde{\epsilon}^{e}_{k'}
    - 
    \tilde{\epsilon}^{e}_{k}
  \right)
  c_{e}^{q,k',k}
}
\nonumber\\
 &&
 + V_{k-k'-q} 
  \left[ 
    f^e_{k-q} f^e_{k'+q}
    \left(1- f^e_{k}\right)
    \left(1- f^e_{k'}\right)
-
    f^e_{k}
    f^e_{k'}
    \left(1 - f^e_{k-q} \right)
    \left(1 - f^e_{k'+q} \right)
  \right]
\nonumber\\
&&
  -V_q 
  \left[ 
    f^e_{k-q} f^e_{k'+q}
    \left(1- f^e_{k}\right)
    \left(1- f^e_{k'}\right)
-
    f^e_{k}
    f^e_{k'}
    \left(1 - f^e_{k-q} \right)
    \left(1 - f^e_{k'+q} \right)
\right]
\nonumber\\
   &&
   +V_q
   \left[
     \left(
       f^e_{k'+q}-f^e_{k'}
     \right)
     \sum_{l}
     c_{{e+X}}^{k-q-l,l,k}
     -
     \left(
       f^e_{k}-f^e_{k-q}
     \right)
     \sum_{l}
     c_{{e+X}}^{q+k'-l,l,k'}
     \right]
\nonumber\\
&&
   -V_{k-k'-q}
   \left[
     \left(
       f^e_{k'+q}-f^e_{k}
     \right)
     \sum_{l}
     c_{{e+X}}^{k-q-l,l,k'}
     - 
     \left(
       f^e_{k'}-f^e_{k-q}
     \right)
     \sum_{l}
     c_{{e+X}}^{q+k'-l,l,k}
   \right]
\nonumber\\
&&
   +\left[
     1-f^e_{k-q}-f^e_{k'+q}
   \right]
   \sum_{l} 
   V_{l-q}
   c_{e}^{l,k',k}
-
   \left[
     1-f^e_{k}-f^e_{k'}
   \right]
   \sum_{l}
   V_{l-q}
   c_{e}^{l,q+k'-l,k-q+l}
\nonumber\\
&&
   -\left[
     f^e_{k}-f^e_{k-q}
   \right]
   \sum_{l} 
   V_{l-k}
   c_{e}^{q,k',l}
+
   \left[
     f^e_{k'+q}-f^e_{k}
   \right]
   \sum_{l} 
   V_{l-k}
   c_{e}^{q-k+l,k',l}
\nonumber\\
&&
  +\left[
     f^e_{k'+q}-f^e_{k'}
   \right]
   \sum_{l} 
   V_{l-k'}
   c_{e}^{q,l,k}
-
   \left[
     f^e_{k'}-f^e_{k-q}
   \right]
   \sum_{l} 
   V_{l-k'}
   c_{e}^{q+k'-l,l,k}
\label{eq:cccc},
\end{eqnarray}
%%%%%%%%%%%%%%%%%%%%%%%
%%%%%%%%%%%%%%%%%%%%%%%
\begin{eqnarray}
\lefteqn{
  \left[
  i \hbar\ddt
  c_{h}^{q,k',k} 
  \right]_{H_{\mrm{kin}}+H_{\mrm{C}}}
=
\left(
    -\tilde{\epsilon}^{h}_{k-q}
    -
    \tilde{\epsilon}^{h}_{k'+q} 
    + 
    \tilde{\epsilon}^{h}_{k'}
    + 
    \tilde{\epsilon}^{h}_{k}
  \right)
  c_{h}^{q,k',k}
}
\nonumber\\
 &&
 + V_{k-k'-q} 
  \left[ 
    \left(1-f^h_{k-q}\right) 
    \left(1-f^h_{k'+q}\right) 
    f^h_{k}
    f^h_{k'}
-
    \left(1 - f^h_{k}\right)
    \left(1-f^h_{k'}\right)
    f^h_{k-q} 
    f^h_{k'+q} 
  \right]
\nonumber\\
&&
  -V_q 
  \left[ 
    \left(1-f^h_{k-q}\right)
    \left(1-f^h_{k'+q}\right)
    f^h_{k} f^h_{k'}
-
    \left(1-f^h_{k}\right)
    \left(1-f^h_{k'}\right)
    f^h_{k-q}f^h_{k'+q}
\right]
\nonumber\\
   &&
   +V_q
   \left[
     \left(f^h_{k'}-
       f^h_{k'+q}
     \right)
     \sum_{l}
     c_{{h+X}}^{q-k+l,k,l}
     -
     \left(
       f^h_{k-q}-f^h_{k}
     \right)
     \sum_{l}
     c_{{h+X}}^{l-q-k',k',l}
     \right]
\nonumber\\
&&
   +V_{k-k'-q}
   \left[
     \left(
       f^h_{k'+q}-f^h_{k}
     \right)
     \sum_{l}
     c_{{h+X}}^{q-k+l,k',l}
     - 
     \left(
       f^h_{k'}-f^h_{k-q}
     \right)
     \sum_{l}
     c_{{e+X}}^{l-q-k',k,l}
   \right]
\nonumber\\
&&
   -\left[
     1-f^h_{k-q}-f^h_{k'+q}
   \right]
   \sum_{l} 
   V_{l-q}
   c_{h}^{l,k',k}
+
   \left[
     1-f^h_{k}-f^h_{k'}
   \right]
   \sum_{l}
   V_{l-q}
   c_{h}^{l,q+k'-l,k-q+l}
\nonumber\\
&&
   +\left[
     f^h_{k}-f^h_{k-q}
   \right]
   \sum_{l} 
   V_{l-k}
   c_{h}^{q,k',l}
-
   \left[
     f^h_{k'+q}-f^h_{k}
   \right]
   \sum_{l} 
   V_{l-k}
   c_{h}^{q-k+l,k',l}
\nonumber\\
&&
  -\left[
     f^h_{k'+q}-f^h_{k'}
   \right]
   \sum_{l} 
   V_{l-k'}
   c_{h}^{q,l,k}
+
   \left[
     f^h_{k'}-f^h_{k-q}
   \right]
   \sum_{l} 
   V_{l-k'}
   c_{h}^{q+k'-l,l,k}
\label{eq:vvvv}.
\end{eqnarray}
%%%%%%%%%%%%%%%%%%%%%%%
%
%%%%%%%%%%%%%%%%%%%%%%%%%%%%%%%%%%%%%%%%%%%%%%%%%%%
\section{Phonon interaction}
\label{app:phonons}
%%%%%%%%%%%%%%%%%%%%%%%%%%%%%%%%%%%%%%%%%%%%%%%%%%%
%
The general phonon interaction Hamiltonian Eq.~(\ref{eq:HP}) is the starting point to
compute the operator equation of motion for a two-point operator
%
%%%%%%%%%%%%%%%%%%%%
%%%%%%%%%%%%%%%%%%%%
\begin{equation}
\left[
i\hbar \ddt a^{\dagger}_{\lambda, k} a_{\lambda, k+q}
\right]_{H_\mrm{P}} = 
	\sum_{p} \mcal{G}_{p} 	\left(	a^{\dagger}_{\lambda,k} a_{\lambda, k+q-p} 
					-
					a^{\dagger}_{\lambda,k+p} a_{\lambda, k+q}
				\right),
\label{eq:basicphon}
\end{equation}
%%%%%%%%%%%%%%%%%%%%
%%%%%%%%%%%%%%%%%%%%
%
where the collective phonon operator 
%
%%%%%%%%%%%%%%%%%%%%
%%%%%%%%%%%%%%%%%%%%
\begin{equation}
	\mcal{G}_{p} =
		\sum_{p_\perp} 
		G_{p,p_\perp} 
	\left( D_{p,p_\perp} + D^{\dagger}_{-p,p_\perp} \right)
\label{eq:Gphon}
\end{equation}
%%%%%%%%%%%%%%%%%%%%
%%%%%%%%%%%%%%%%%%%%
%
has been defined. With the help of this operator equation, we can now generate all
other equations of motion. The simplest case is the equation for carrier densities.
For electrons, we obtain
%
%%%%%%%%%%%%%%%%%%%%
%%%%%%%%%%%%%%%%%%%%
\begin{equation}
\left[
i\hbar \ddt f^{e}_{k}
\right]_{H_\mrm{P}} = 
	\sum_{p}
		\Delta\ev{\mcal{G}_{p} a^{\dagger}_{c,k} a_{c,k-p} }
					- \mbox{h.c.}
\label{eq:fephon1}
\end{equation}
%%%%%%%%%%%%%%%%%%%%
%%%%%%%%%%%%%%%%%%%%
%
In order to compute the right-hand side of Eq.~(\ref{eq:fephon1}), we have
to establish the equations of motion for the phonon assisted terms.
In the case of the phonon assisted electron density, this gives
%
%%%%%%%%%%%%%%%%%%%%
%%%%%%%%%%%%%%%%%%%%
\begin{eqnarray}
\left[
i\hbar \ddt \Delta\ev{D_{p,p_\perp} a^{\dagger}_{c,k} a_{c,k-p}}
\right]_{H_\mrm{P}} & = &
	G_{p,p_\perp} f^{e}_{k-p} (1-f^{e}_{k})
\nonumber\\
&& + \Delta\ev{D_{p,p_\perp} \mcal{G}_{-p}} (f^{e}_{k} - f^{e}_{k-p}) 
\nonumber\\
&& + G_{p,p_\perp} \sum_{l} \left( c_{e}^{p,l,k} - c_{X}^{k-p-l,l,k} \right) 
\label{eq:Dfephon1},
\end{eqnarray}
%%%%%%%%%%%%%%%%%%%%
%%%%%%%%%%%%%%%%%%%%
%
where we have already left out all terms proportional to the coherent transition
amplitudes $P_k=\ev{v^{\dagger}_{k} c_{k}}$. In cases with coherent excitations,
those terms can easily be included. But since in this case the Coulomb scattering
is usually dominant, the main application of the microscopic phonon scattering
lies in the incoherent regime. In order to compute Eq.~(\ref{eq:fephon1}), we
solve Eq.~(\ref{eq:Dfephon1}) in Markov approximation. In principle, the result of
a Markov approximation is dependent on the basis which we use. In general, however,
the difference is not crucial if the resulting terms are summed over as is the
case in Eq.~(\ref{eq:fephon1}). For the computations, we have therefore included
only terms proportional to the square of the phonon matrix elements. No
Coulomb or light matter interaction was included in the equation of phonon assisted
densities. The Markov approximation was done in the one-particle basis of Bloch
electrons and holes. We numerically confirmed that our result is independent of the
precise choice of the basis. Additionally, we assume a reservoir of phonons, i.e.,
we neglect coherent phonons and set
$\ev{D^\dagger_{p,p_\perp} D_{p,p_\perp}}=
\left( \exp(E_{p,p_\perp}/(kT)) - 1 \right)^{-1}$ to 
the Bose-Einstein distribution at the corresponding energy $E_{p,p_\perp}$ given by
the phonon dispersion. After those approximations, we obtain Eqs.~(\ref{eq:ddtfephon2})
and~(\ref{eq:ddtfhphon2}) for the carrier densities.

We have seen in Sec.~\ref{sec:eomHP} that Eq.~(\ref{eq:Dfephon1}) and its counterpart
for the phonon assisted hole densities suffice to compute the doublet
contributions to the correlation equations. In addition, phonon assisted
correlations of the form $\Delta\ev{D a^\dagger a^\dagger aa}$ were needed in
order to provide a true dephasing mechanism for excitons and carrier-carrier
correlations on the triplet level. More specifically, the phonon contribution to
the equation of motion for excitonic correlations is
%
%%%%%%%%%%%%%%%%%%%%
%%%%%%%%%%%%%%%%%%%%
\begin{eqnarray}
\lefteqn{
i\hbar \ddt \Delta\ev{a^\dagger_{c,k} a^\dagger_{v,k'} a_{c,k'+q} a_{v,k-q}} =}
\nonumber\\
&&
-
\Delta\ev{{\cal G}^\dagger_{k'+q-k} a^\dagger_{c,k} a_{c,k'+q}} 
\left( f^{h}_{k-q} - f^{h}_{k'} \right)
-
\Delta\ev{{\cal G}_{k'+q-k} a^\dagger_{v,k'} a_{v,k-q}} 
\left( f^{e}_{k} - f^{e}_{k'+q} \right)
\nonumber\\
&&
+
\sum_{p}\Delta
\ev{{\cal G}^{\dagger}_{p} a^\dagger_{c,k} a^\dagger_{v,k'} a_{c,k'+q} a_{v,k-q+p}}
-
\sum_{p}\Delta
\ev{{\cal G}^{\dagger}_{p} a^\dagger_{c,k-p} a^\dagger_{v,k'} a_{c,k'+q} a_{v,k-q}}
\nonumber\\
&&
+
\sum_{p}\Delta
\ev{{\cal G}^{\dagger}_{p} a^\dagger_{c,k} a^\dagger_{v,k'} a_{c,k'+q+p} a_{v,k-q}}
-
\sum_{p}\Delta
\ev{{\cal G}^{\dagger}_{p} a^\dagger_{c,k} a^\dagger_{v,k'-p} a_{c,k'+q} a_{v,k-q}}
\label{eq:ddtcXphon1}
\end{eqnarray}
%%%%%%%%%%%%%%%%%%%%
%%%%%%%%%%%%%%%%%%%%
%
and similar equations exist also for the carrier-carrier correlations.
The first row involves the phonon assisted densities solved already in
Eq.~(\ref{eq:Dfephon1}). The remaining phonon assisted correlations are
solved via
%
%%%%%%%%%%%%%%%%%%%%
%%%%%%%%%%%%%%%%%%%%
\begin{eqnarray}
\lefteqn{
i\hbar \ddt \Delta\ev{D_{p,p_\perp} a^\dagger_{c,k} a^\dagger_{v,k'} a_{c,k'+q} a_{v,k-q-p} }
= 
}\nonumber\\
&& 
\phantom{+}
G_{p,p_\perp} \left( N^{\text{PH}}_{p,p_\perp} + f^{h}_{k-q-p}  \right)
\Delta\ev{a^\dagger_{c,k} a^\dagger_{v,k'} a_{c,k'+q} a_{v,k-q}}
\nonumber\\
&&
+
G_{p,p_\perp} \left( N^{\text{PH}}_{p,p_\perp} + 1 - f^{e}_{k'+q}  \right)
\Delta\ev{a^\dagger_{c,k} a^\dagger_{v,k'} a_{c,k'+q+p} a_{v,k-q-p}}
\nonumber\\
&&
-
G_{p,p_\perp} \left( N^{\text{PH}}_{p,p_\perp} + 1 - f^{h}_{k'}  \right)
\Delta\ev{a^\dagger_{c,k} a^\dagger_{v,k'-p} a_{c,k'+q} a_{v,k-q-p}}
\nonumber\\
&&
-
G_{p,p_\perp} \left( N^{\text{PH}}_{p,p_\perp} + f^{e}_{k}  \right)
\Delta\ev{c^\dagger_{a,k-p} a^\dagger_{v,k'} a_{c,k'+q} a_{v,k-q-p}}
\label{eq:ddtDcvcvphon}
\end{eqnarray}
%%%%%%%%%%%%%%%%%%%%
%%%%%%%%%%%%%%%%%%%%
%
which together with similar terms for phonon-assisted carrier-carrier correlations
can be used to build up the complete phonon interaction contributions to the 
correlation equations. Again, we have kept only incoherent terms which enter the
correlation equations proportional to the square of the phonon matrix element.
Equation~(\ref{eq:ddtDcvcvphon}) is solved in Markov approximation assuming a
phonon bath characterized by the lattice temperature. This way, we obtain 
Eq.~(\ref{eq:ddtcXphon6p}) and the corresponding equations
%
%%%%%%%%%%%%%%%%%%%%
%%%%%%%%%%%%%%%%%%%%
\begin{eqnarray}
\lefteqn{
\left[
  i \hbar\ddt
  c_{e}^{q,k',k} 
\right]_{H_{\mrm{P}},\mrm{T}}
=
\sum_{p}
\left(
\tilde{\gamma}^{e}_{k-q,p} + \tilde{\gamma}^{e}_{k'+q,p}
-
(\tilde{\gamma}^{e}_{k',p})^* - (\tilde{\gamma}^{e}_{k,p})^*
\right)
c_e^{q,k',k}
}
\nonumber\\
&&
-
\sum_{p}
\left(
\gamma^{e}_{k,k-p} - (\gamma^{e}_{k-q,k-p})^*
\right)
c_e^{q,k',p}
-
\sum_{p}
\left(
\gamma^{e}_{k',k'-p} - (\gamma^{e}_{k'+q,k'-p})^*
\right)
c_e^{q,p,k}
\nonumber\\
&&
-
\sum_{p}
\left(
(\gamma^{e}_{k-q,p-q})^* + (\gamma^{e}_{k'+q,q-p})^*
\right)
c_e^{p,k',k}
+
\sum_{p}
\left(
\gamma^{e}_{k',p-q} + \gamma^{e}_{k,q-p}
\right)
(c_e^{p,k-q,k'+q})^*
\nonumber\\
&&
-
\sum_{p}
\left(
\gamma^{e}_{k',p-q} - (\gamma^{e}_{k-q,p-q})^*
\right)
c_e^{p,k'+q-p,k}
-
\sum_{p}
\left(
\gamma^{e}_{k,q-p} - (\gamma^{e}_{k'+q,q-p})^*
\right)
c_e^{p,k',k-q+p}
\label{eq:ddtcephon6p}
\end{eqnarray}
%%%%%%%%%%%%%%%%%%%%
%%%%%%%%%%%%%%%%%%%%
%
and
%
%%%%%%%%%%%%%%%%%%%%
%%%%%%%%%%%%%%%%%%%%
\begin{eqnarray}
\lefteqn{
\left[
  i \hbar\ddt
  c_{h}^{q,k',k}
\right]_{H_{\mrm{P}},\mrm{T}}
=
\sum_{p}
\left(
\tilde{\gamma}^{h}_{k-q,p} + \tilde{\gamma}^{h}_{k'+q,p}
-
(\tilde{\gamma}^{h}_{k',p})^* - (\tilde{\gamma}^{h}_{k,p})^*
\right)
c_h^{q,k',k}
}
\nonumber\\
&&
-
\sum_{p}
\left(
\gamma^{h}_{k,k-p} - (\gamma^{h}_{k-q,k-p})^*
\right)
c_h^{q,k',p}
-
\sum_{p}
\left(
\gamma^{h}_{k',k'-p} - (\gamma^{h}_{k'+q,k'-p})^*
\right)
c_h^{q,p,k}
\nonumber\\
&&
-
\sum_{p}
\left(
(\gamma^{h}_{k-q,p-q})^* + (\gamma^{h}_{k'+q,q-p})^*
\right)
c_h^{p,k',k}
+
\sum_{p}
\left(
\gamma^{h}_{k',p-q} + \gamma^{h}_{k,q-p}
\right)
(c_h^{p,k-q,k'+q})^*
\nonumber\\
&&
-
\sum_{p}
\left(
\gamma^{h}_{k',p-q} - (\gamma^{h}_{k-q,p-q})^*
\right)
c_h^{p,k'+q-p,k}
-
\sum_{p}
\left(
\gamma^{h}_{k,q-p} - (\gamma^{h}_{k'+q,q-p})^*
\right)
c_h^{p,k',k-q+p}
\label{eq:ddtchphon6p}
\end{eqnarray}
%%%%%%%%%%%%%%%%%%%%
%%%%%%%%%%%%%%%%%%%%
%
where the terms $\gamma$ and $\tilde{\gamma}$ are identical to those defined
in Eqs.~(\ref{eq:definegammae})--(\ref{eq:definegammatildeh}).

%
%%%%%%%%%%%%%%%%%%%%%%%%%%%%%%%%%%%%%%%%%%%%%%%%%%%%%
\end{document}